\documentclass{article}
\usepackage{amsmath,amssymb}
\usepackage{jheppub}
\makeatletter
\newcommand\fake@math{}
\def\fake@math#1\){{#1}}
\makeatother

\usepackage{graphicx}

\newcommand{\Ber}{\mathop{\mathrm{Ber}}\nolimits}
\newcommand{\Res}{\mathop{\mathrm{Res}}\nolimits}
\newcommand{\sgn}{\mathop{\mathrm{sgn}}\nolimits}

\numberwithin{equation}{section}

\begin{document}
\title{\(\mathcal N=2\) SYK model in the superspace formalism}
\author[a]{Ksenia Bulycheva}
\abstract{We use superspace methods to study an SYK-like model with $\mathcal N=2$ supersymmetry in one dimension, and an analog of this model in two dimensions.
   We find the four-point function as an expansion in the basis of eigenfunctions of the Casimir of $su(1,1|1)$.
We also find retarded kernels and Lyapunov exponents for both cases.}

\affiliation[a]{Department of Physics, Princeton University, Princeton, NJ 08544}
\emailAdd{kseniab@princeton.edu}


\maketitle


\section{Introduction}

Since its introduction \cite{Sachdev:1992fk}, \cite{Kitaev:2015}, the SYK model has been generalized in many ways.
It has been endowed with extra global symmetry \cite{Gross:2016kjj}, \cite{Yoon:2017nig}, supersymmetry \cite{Fu:2016vas}, \cite{Murugan:2017eto}, \cite{Peng:2017spg}, it has been studied as a tensor model with non-random coupling \cite{Witten:2016iux}, \cite{Klebanov:2016xxf}, also with added supersymmetry \cite{Peng:2016mxj}.
In this paper, we study an $\mathcal N=2$ supersymmetric version of the model, and then generalize further to a two-dimensional theory.

The study of the SYK models with extra symmetries largely follows the scheme developed in \cite{Maldacena:2016hyu}.
The two-point function of the model is found from Schwinger--Dyson equations, following immediately from the Lagrangian.
The four-point function can be found directly from summing ladder diagrams, but this is rather tricky; instead, the four-point function is expanded in the basis of eigenfunctions of the Casimir of the corresponding superconformal group.
The four-point function contains information about operator content of the theory; also, by means of the out-of-time ordered four-point functions we can find the chaos exponent, which is one of the main attractive features of this model.
This is the scheme we are following in this paper as well.

Supersymmetric generalizations \cite{Fu:2016vas} of the model are interesting for several reasons. 
First, they allow us to study two-dimensional versions of the SYK model.
In two dimensions, fermions have scaling dimension 1/2, so a relevant interaction cannot be constructed from fermions only.
In contrast, two-dimensional scalars have scaling dimension zero, but a bosonic random potential can have negative directions.
To cure that, one can consider a supersymmetric two-dimensional model of scalar superfields with a random superpotential.
In an $\mathcal N=2$ supersymmetric SYK model, we consider chiral superfields with a random holomorphic superpotential.
%

A two-dimensional $\mathcal N=2$ model with a (quasi)homogeneous holomorphic superpotential is generally assumed to flow to a conformal fixed point \cite{Vafa:1988uu}.
SYK models with less supersymmetry are conformal in the infrared limit at large $N$, but one might expect that $1/N$ corrections induce a ``slow'' RG flow and drive the system away from the conformal point.
Such corrections are hard to study and little is known about them to date.
In contrast, we expect the $\mathcal N=2$ model to flow to a true conformal point, which we can conveniently study in the large $N$ limit with the methods designed for the usual non-supersymmetric SYK.

Although we don't discuss this question in the paper, we notice that constructing a gravity dual of SYK is a challenging task.
The similarities between SYK and $AdS_2$ gravity has already been noticed in the early papers on the subject 
\cite{Almheiri:2014cka}, \cite{Maldacena:2016hyu}, \cite{Jensen:2016pah}, \cite{Maldacena:2016upp}, \cite{Engelsoy:2016xyb}, however the full understanding of a gravity dual is still missing, except for some particular cases as in \cite{Das:2017pif}.
We hope that adding extra supersymmetry might shed some light on this question as well.

The $\mathcal N=2$ SYK model has already been studied in \cite{Fu:2016vas} and \cite{Peng:2017spg}.
In this paper, we develop the approach of \cite{Fu:2016vas} and work in superspace with chiral and anti-chiral fields.
The $\mathcal N=2$ supersymmetry allows complex superfields, and therefore we have to consider four-point functions with different parity under exchange of incoming particles.
In this respect, it is very similar to the SYK model with complex fermions we have studied in \cite{Bulycheva:2017uqj}.
Also, the $SU(1,1|1)$ superconformal group is large enough to restrict the odd coordinates in the chiral--anti-chiral four-point function to zero.
We see that the eigenfunctions of the Casimir turn out to be purely bosonic, and in fact linear combinations of the $\mathcal N=0$ eigenfunctions.

This paper is a logical continuation of \cite{Bulycheva:2017uqj} and relies heavily on the machinery developed in \cite{Murugan:2017eto}.
We also compare some of our results against \cite{Fu:2016vas} and \cite{Peng:2017spg} and find them in agreement.

The structure of this paper is the following.
In Section \ref{sec:superspace} we introduce $\mathcal N=2$ superspace and superfields.
In Section \ref{sec:2pt} we write the Lagrangian of the model and discuss the conformal two-point function found from the Schwinger--Dyson equation.
In Section \ref{sec:4pt} we discuss the two-particle superconformal Casimir and write its eigenfunctions in the shadow representation. 
Then we find the norm of the eigenfunctions and the eigenvalues of the SYK kernel acting on them.
It allows us to write the full four-point function as a series.
In Section \ref{sec:retarded} we find the retarded kernel and compute the Lyapunov exponent corresponding to the superconformal charge multiplet which turns out to be maximal.
Finally, in Section \ref{sec:2d} we generalize some of our results to two dimensions.

{\bf Acknowledgments.} The author is grateful to Edward Witten for suggesting the problem and discussions on the subject, and also to Douglas Stanford for useful conversations.
The author appreciates the hospitality of the Simons Center for Geometry and Physics where a part of this work has been done.

\section{\(\mathcal N=2\) superspace and superfields}
\label{sec:superspace}

We study the $\mathcal N=2$ model at large $N$ in the strong coupling limit. 
The model flows to a theory which possesses the full $SU(1,1|1)$ superconformal symmetry.
To study the correlators, it is convenient to work in the one-dimensional $\mathcal N=2$ superspace 
(with Euclidean signature), parameterized by:

\begin{equation}
    \left(\tau, \theta, \bar{\theta}\right).
    \label{super_coord}
\end{equation}
In what follows, we will often substitute this set of coordinates with a single number representing the index of the supercoordinate, for example:
\begin{equation}
    \Phi\left( 1 \right)\equiv \Phi\left( \tau_1, \theta_1, \bar{\theta}_1 \right).
    \label{index_def}
\end{equation}
The $SU(1,1|1)$ group has four bosonic and four fermionic coordinates.
It is generated by super-translations:
\begin{equation}
    \tau \to \tau+ \epsilon + \theta \bar{\eta} + \bar{\theta} \eta, \qquad \theta \to \theta+\eta, \qquad \bar{\theta} \to \bar{\theta} + \bar{\eta},
    \label{super_trans}
\end{equation}
inversions:
\begin{equation}
    \tau \to -\frac 1\tau , \qquad \theta \to \frac{\theta}{\tau}, \qquad \bar{\theta} \to \frac{\bar{\theta}}{\tau},
    \label{inversion}
\end{equation}
and the $R$--symmetry transformation:

\begin{equation}
    \theta \to e^{i\alpha} \theta, \qquad \bar{\theta} \to e^{-i\alpha} \bar{\theta}.
    \label{R_action}
\end{equation}
In Appendix \ref{sec:app_Casimir}, we write down the generators of the $su(1,1|1)$ superconformal group as differential operators in the superspace.

The correlators in a CFT have to be conformally covariant.
In particular, they have to be invariant under translations, which in non-supersymmetric theory makes them depend only on differences of coordinates:

\begin{equation}
    \tau_{12}=\tau_1-\tau_2.
    \label{diff_n=0}
\end{equation}

In the supersymmetric case, this condition gets more restrictive and correlation functions are invariant under super-translations, together with $R$-symmetry.
We can write two combinations of super-coordinates with conformal weight $-1$ which satisfy these restrictions:

\begin{equation}
    \Delta_{12}\equiv\tau_1-\tau_2-\theta_1 \bar{\theta}_2 - \bar{\theta}_1 \theta_2,  \qquad \lambda_{12}\equiv \left( \theta_1-\theta_2 \right)\left( \bar{\theta}_1-\bar{\theta}_2 \right).
    \label{delta_lambda_def}
\end{equation}
These two combinations have different symmetry under $1 \leftrightarrow 2$ permutation:

\begin{equation}
    \Delta_{12}=-\Delta_{21}, \qquad \lambda_{12}=\lambda_{21}.
    \label{delta_lambda_symm}
\end{equation}
The correlators should be functions of $\Delta$, $\lambda$.
In fact, we can restrict them even further using chirality constraint.
The complex fermions and bosons in the model can be arranged into chiral superfields $\Psi, \bar{\Psi}$ satisfying:

\begin{equation}
    \bar{D}\Psi=0, \qquad D\bar{\Psi}=0,
    \label{Psi_chiral}
\end{equation}
where $D, \bar{D}$ are super-derivatives:

\begin{equation}
    D\equiv \frac{\partial}{\partial \theta}+\bar{\theta} \frac{\partial}{\partial \tau}, \qquad \bar{D} \equiv \frac{\partial}{\partial \bar{\theta}}+ \theta \frac{\partial }{\partial \tau}.
    \label{D_def}
\end{equation}

Correlators of chiral (anti-chiral) fields are also chiral (or anti-chiral):
\begin{equation}
    \bar{D}_1 \langle \Psi\left( 1 \right) \dots \rangle=0.
    \label{D<>}
\end{equation}
Therefore they should depend on a chiral (anti-chiral) combination of the super-translation invariants $\Delta, \lambda$.
Let us find a linear combination annihilated by $D$:

\begin{equation}
    \langle 12 \rangle = \Delta_{12} - \lambda_{12} = \tau_1-\tau_2-2\bar{\theta}_1 \theta_2-\theta_1 \bar{\theta}_1-\theta_2 \bar{\theta}_2.
    \label{<>_def}
\end{equation}
This choice is unique, and the nice thing about this invariant combination is that it is both chiral in the first coordinate and anti-chiral in the second one:

\begin{equation}
    D_1 \langle 12 \rangle =\bar{D}_2 \langle 12 \rangle=0.
    \label{D_<>}
\end{equation}
It makes writing the correlators particularly easy.
For example, the two-point function can depend only on the $\langle 12 \rangle$ combination:

\begin{equation}
    \mathcal G\left( 1|2 \right)\equiv\mathcal G\left( \tau_1, \theta_1, \bar{\theta}_1|  \tau_2, \theta_2, \bar{\theta}_2 \right)\equiv\langle \bar{\Psi}\left( \tau_1, \theta_1, \bar{\theta}_1 \right) \Psi \left( \tau_2, \theta_2, \bar{\theta}_2 \right)\rangle=\mathcal G\left( \langle 12 \rangle \right).
    \label{G_<>}
\end{equation}
Likewise, the three-point function combining a chiral and an antichiral fields with some superfield $V$ is a function of three invariants:
\begin{equation}
    \langle \bar{\Psi}\left( 1 \right)\Psi \left( 2 \right) V\left( 0 \right)\rangle=f\left( \langle 12 \rangle, \langle 10 \rangle, \langle 02 \rangle \right).
    \label{3pt_<>}
\end{equation}
To make this three-point function non-trivial, the $R$-charge of the $\mathcal  V$ operator has to vanish.
It means in particular that $\mathcal V$ cannot be a chiral or an anti-chiral superfield.

In what follows we write all the correlation functions in terms of the $\langle ij \rangle$ invariants.
This makes the correlators manifestly supersymmetric.
Using the superconformal group sometimes helps us fix most of the odd variables, so that the results can written as functions of purely bosonic variables; however, the odd variables are generally easy to reinstall back.
This can be used to find the correlation functions of the component fields, although we are not following this approach here.

\section{Two-point function}
\label{sec:2pt}
We are studying correlators of chiral superfields $\Psi, \bar{\Psi}$, written in the $\mathcal N=2$ superspace. 
The Lagrangian of the model consists of a kinetic $F$-term and a holomorphic superpotential:
\begin{equation}
    \mathcal L=\int d\bar{\theta} d\tau \bar{\Psi}_iD \Psi_i + i^{\frac{\hat{q}-1}{2}}\int d\theta d\tau  C_{i_1 i_2 \dots i_{\hat{q}}}\Psi_{i_1} \dots \Psi_{i_{\hat{q}}}+ i^{\frac{\hat{q}-1}{2}}\int d\bar{\theta} d\tau  \bar{C}_{i_1 i_2 \dots i_{\hat{q}}}\bar{\Psi}_{i_1} \dots \bar{\Psi}_{i_{\hat{q}}},
    \label{Lagr_1d}
\end{equation}
with the random Gaussian coupling:
\begin{equation}
    \langle C_{i_1 \dots i_{\hat{q}}}\bar{C}_{i_1 \dots i_{\hat{q}}}\rangle = \left( \hat{q}-1 \right)!\frac{J}{N^{\hat{q}-1}},
    \label{CC}
\end{equation}
$\hat{q}$ being an arbitrary odd integer.

$\Psi$ is a chiral superfield annihilated by $\bar{D}$, so in components it reads as:
\begin{equation}
    \Psi=\psi\left( \tau+ \theta \bar{\theta}\right)+\theta b.
    \label{Psi_comp}
\end{equation}
$\psi$, $b$ are complex fermion and scalar. 
From the Lagrangian (\ref{Lagr_1d}) we see that the scalar field is non-dynamical.
We can integrate it out and find that the effective Lagrangian has the schematic form:

\begin{equation}
    \mathcal L_{eff}=\int d\tau \left( \bar{\psi}\partial_\tau\psi+ C\bar{C} \bar{\psi}^{q/2} \psi^{q/2} \right),
    \label{Lagr_1d_nob}
\end{equation}
with $q=2\hat{q}-2$.
It is very similar to the Lagrangian of the non-supersymmetric SYK model for complex fermions (although the coupling $C\bar{C}$ has different structure), so we can expect the story to be reminiscent of the non-supersymmetric case.

\begin{figure}
   \centering
   \includegraphics[width=.7\textwidth]{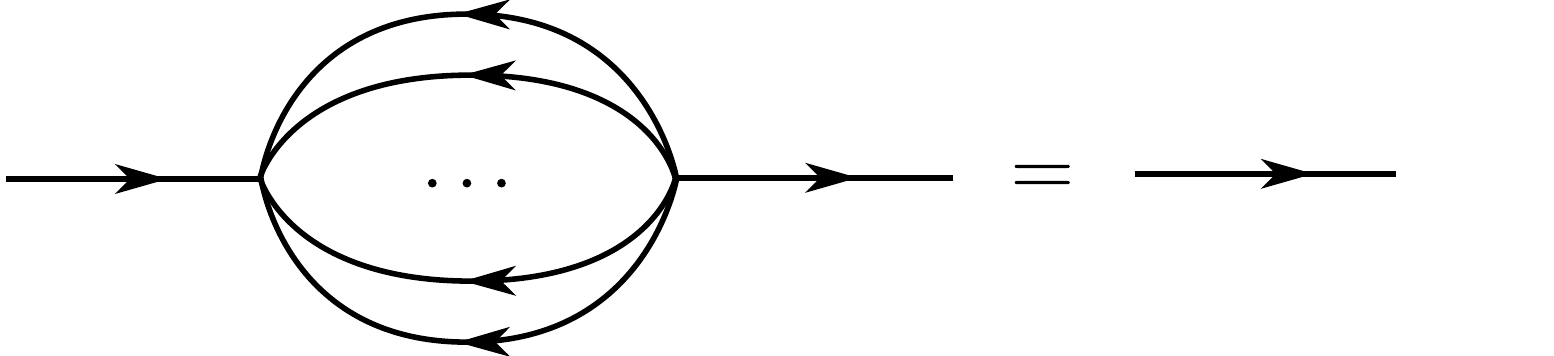}
   \caption{Schwinger--Dyson equation for the two-point function. The melonic part contains an even number of propagators.}
   \label{fig:SD}
\end{figure}

Now we can find the conformal two-point function of the superfield.
Keeping in mind (\ref{G_<>}), we look for the propagator of the form:

\begin{equation}
    \mathcal G \left(1| 2 \right)=\mathcal G\left( \langle 12\rangle \right)=b\frac{\sgn \left( \left \langle 12 \right \rangle  \right)}{|\langle 12\rangle|^{2\Delta}},
    \label{G_form}
\end{equation}
where $\langle 12\rangle$ is the invariant defined in (\ref{<>_def}). 
The propagator has to satisfy the Schwinger--Dyson equation.
We can read it off the Lagrangian (\ref{Lagr_1d}). 
Neglecting the $D \mathcal G$ term, we find the equation to be (see fig. \ref{fig:SD}):

\begin{equation}
    \int d\tau_1 d\theta_1 J \mathcal G\left( \langle 01 \rangle \right) \mathcal G \left( \langle 21 \rangle \right)^{\hat{q}-1}=\left(\bar{\theta}_0-\bar{\theta}_2  \right)\delta\left( \langle 02 \rangle \right).
    \label{SD_1d}
\end{equation}
The delta-function has to be chiral in the first coordinate, hence it depends only on $\langle 02\rangle$ (and therefore is anti-chiral in the second coordinate). 
The value of $\Delta$ follows from dimensional considerations:
\begin{equation}
    2\Delta \hat{q}=1.
    \label{Delta_q}
\end{equation}
To find $b$ and check the ansatz (\ref{G_form}), we integrate over odd variables in the Schwinger--Dyson equation and then make a one-dimensional Fourier transformation, using the integral:
\begin{equation}
    \int d\tau \frac{1}{|\tau|^{2\Delta}}e^{i\omega \tau}=\sqrt{\frac{2}{\pi}}|\omega|^{-1+2\Delta}\Gamma\left( 1-2\Delta \right) \sin \pi \Delta. 
    \label{1d_fourier}
\end{equation}
Then the $b$ constant is fixed to:
\begin{equation}
    4\pi Jb^{\hat{q}} =\tan \pi \Delta.
    \label{b_1d}
\end{equation}

The four-point function in the model can also be found from an integral equation. 
To solve it, we use the fact that the integral kernel commutes with Casimir of the conformal group, and therefore they have a common basis of eigenfunctions.
In the next Section, we find eigenfunctions of the Casimir and expand the four-point function in this basis.

\section{Four-point function}
\label{sec:4pt}
We are looking for a four-point function with two chiral and two anti-chiral fermions:

\begin{equation}
    W\left( 1,2|3,4 \right)\equiv\langle \bar{\Psi} \left( 1 \right) {\Psi} \left(2 \right) \bar{\Psi} \left( 3 \right) {\Psi} \left( 4\right) \rangle.
    \label{4pt_chiral}
\end{equation}

After dividing by propagators, this four-point function becomes invariant under the superconformal group:

\begin{equation}
    \mathcal W \equiv \frac{W}{\mathcal G \left( \langle 12 \rangle \right) \mathcal G \left( \langle 34 \rangle \right)}.
    \label{W_inv_def}
\end{equation}
It means that $\mathcal W$ can depend only on the cross-ratio of the coordinates. 
Unlike the non-supersymmetric and $\mathcal N=1$ supersymmetric cases, there is only one cross-ratio consistent with chirality, namely:

\begin{equation}
    \chi\equiv \frac{\langle 12 \rangle \langle 34 \rangle}{\langle 14 \rangle \langle 32 \rangle}.
    \label{chi_def}
\end{equation}
There is no nilpotent invariant as in the $\mathcal N=1$ case either.

We can use the superconformal symmetry to fix the coordinates conveniently.
There are four bosonic generators, one of which generates the translation symmetry, and four fermionic ones.
We can use the fermionic generators to fix four out of eight odd coordinates.
Looking at the structure of the invariant (\ref{<>_def}), we see that if we fix $\theta=0$ for the chiral and $\bar{\theta}=0$ for the antichiral fields:
\begin{equation}
    \theta_2=\theta_4=0, \qquad \bar{\theta}_1=\bar{\theta}_3=0,
    \label{theta_fix}
\end{equation}
the cross-ratio reduces to the conventional bosonic cross-ratio:

\begin{equation}
    \chi=\frac{\tau_{12}\tau_{34}}{\tau_{14}\tau_{32}}.
    \label{chi_bos}
\end{equation}
Next we can use the bosonic conformal subgroup to fix three out of four coordinates in the standard way:
\begin{equation}
    \tau_1=\chi, \qquad \tau_2 = 0, \qquad \tau_3=1, \qquad \tau_4=\infty.
    \label{tau_fix}
\end{equation}

This implies that the conformal four-point function is a purely bosonic function and does not depend on odd coordinates, unlike the $\mathcal N=1$ four-point function \cite{Murugan:2017eto}:
\begin{equation}
   \mathcal W = \mathcal W\left( \chi \right).
   \label{W(chi)}
\end{equation}
This also means that the Casimir operator as a differential operator acts only on even coordinates. 
We see in what follows that it is closely related to the Casimir of the non-supersymmetric model.

\subsection{Casimir of \(su(1,1|1)\)}

The most general four-point function can be expanded in the basis of eigenfunctions of the two-particle superconformal Casimir.
We present our convention for the generators and the Casimir of the $su(1,1|1)$ algebra in the Appendix \ref{sec:app_Casimir}.
Conjugating with the two-point functions, we can write the Casimir in terms of the cross-ratio:

\begin{equation}
    C_{1+2} \left( \frac{1}{\langle 12 \rangle^{2\Delta}} \frac{1}{\langle 34 \rangle^{2\Delta}} W(1,2|3,4)\right)=\frac{1}{\langle 12 \rangle^{2\Delta}} \frac{1}{\langle 34 \rangle^{2\Delta}}\mathcal C (\chi)\mathcal W\left( \chi \right),
    \label{C_acts}
\end{equation}
where the conformally-invariant Casimir $\mathcal C\left( \chi \right)$ is a second-order differential operator:
\begin{equation}
    \mathcal C\left( \chi \right)\equiv\chi^2\left( 1-\chi \right)\partial_\chi^2+\chi\left( 1-\chi \right)\partial_\chi.
    \label{C_def}
\end{equation}
This operator is diagonalized by functions $\varphi_h$:

\begin{equation}
    \mathcal C \varphi_h\left( \chi \right)=h^2 \varphi_h\left( \chi \right),
    \label{C_phi}
\end{equation}
which for $\chi<1$ can be expressed in terms of a hypergeometric function:
\begin{equation}
    \varphi_h\left( \chi \right)\equiv \chi^h B\left( h,h \right)\,_2F_1\left( h,h;1+2h;\chi \right), \qquad \chi<1.
    \label{phi_def}
\end{equation}
Notice that the equation (\ref{C_phi}) is symmetric under $h \leftrightarrow -h$, so the basis of the Casimir is spanned by $\varphi_h(\chi)$ and $\varphi_{-h}\left( \chi \right)$.

The Casimir of the $sl(2)$ algebra is very similar to $\mathcal C(\chi)$:
\begin{equation}
    \mathcal C_{\mathcal N=0}=\chi^2\left( 1-\chi \right)\partial_\chi^2-\chi^2\partial_\chi=\mathcal C_{\mathcal N=2}-\chi \partial_\chi,
    \label{C0_C2}
\end{equation}
and the eigenfunctions of the $\mathcal N=0$ and $\mathcal N=2$ SYK models are closely related too. 
If we denote the eigenfunction of the non-supersymmetric model as $F_h(\chi)$:
\begin{equation}
    C_{\mathcal N=0}F_h(\chi)=h\left( h-1 \right)F_h\left( \chi \right), \qquad F_h\left( \chi \right)\equiv B(h,h)\chi^h \,_2F_1\left( h,h;2h; \chi\right) \qquad \text{for } \chi<1,
    \label{C_Fh}
\end{equation}
then the eigenfunction of the $\mathcal N=2$ model $\varphi_h$ is a linear combination:

\begin{equation}
    \varphi_h\left( \chi \right)=F_h\left( \chi \right)-F_{h+1}\left( \chi \right).
    \label{phi_Fh}
\end{equation}
For a proof of this relation see Appendix \ref{sec:app_Fh_phi}.

Knowing the basis of the Casimir, we can fix the  properties of the four-point function  under discrete symmetries (exchange of two fermions) and then find it as a linear combination of $\varphi_h, \varphi_{-h}$.
But we find it advantageous to use the shadow formalism to derive an alternative basis of eigenfunctions, which would already have the desired symmetries by construction.

\subsection{Shadow formalism}

Using the shadow prescription, we treat the fields at the points 1 and 2 as living in a different CFT than the fields at the points 3 and 4. 
Then the four-point function is just a product of independent two-point functions:
\begin{equation}
    W= \mathcal G \left( \langle 12 \rangle \right) \mathcal G \left( \langle 34 \rangle \right)+O(\varepsilon).
    \label{W_no_shadow}
\end{equation}
To find the four-point function, we add a fictitious term to the Lagrangian, which introduces a small coupling between these two CFTs:
\begin{equation}
    \varepsilon\int d\tau_0 d^2\theta_0  V_h \left( \tau_0, \theta_0, \bar{\theta}_0 \right) V'_{-h} \left( \tau_0, \theta_0, \bar{\theta}_0 \right). 
    \label{shadow_int}
\end{equation}
Here $V_h, V'_{-h}$ are fictitious bosonic operators with dimensions adding up to zero, so that the whole integral is dimensionless.

To the first order in $\varepsilon$, this interaction adds to the four-point function an integral of a product of two three-point functions:

\begin{equation}
    W= \mathcal G \left( \langle 12 \rangle \right) \mathcal G \left( \langle 34 \rangle \right)+ \sum_h\varepsilon \int d\tau_0 d^2\theta_0 \langle \bar{\Psi} \left( 1 \right)\Psi\left( 2 \right) V_{h}\left( 0 \right)\rangle \langle \bar{\Psi}\left( 3 \right) \Psi\left( 4 \right) V'_{-h}\left( 0 \right)\rangle+O\left( \varepsilon^2 \right).
    \label{4pt_shadow}
\end{equation}

Now we have to fix the form of chiral-antichiral three-point function. 
In one dimension, a three-point function with two complex fermions can be either odd or even under exchange of those fermions.
Generically it is a linear combination:

\begin{equation}
    \langle \bar{\Psi}\left(1 \right) \Psi \left( 2 \right) V_h \left(0 \right)\rangle =A f_h^A\left( 1,2,0 \right)+S f_h^S\left( 1,2,0 \right).
    \label{3pt_general}
\end{equation}
where the form of the three-point functions is fixed by chirality:
\begin{eqnarray}
    \label{fA_def}
    f_h^A\left( 1,2,0 \right)&=& \frac{\sgn \left( \left \langle 12 \right \rangle  \right)}{ \left|\langle 12\rangle\right|^{2\Delta-h}\left|\langle 10\rangle\right|^{h}\left|\langle 02\rangle\right|^{h}},\\
    \label{fS_def}
    f_h^S\left( 1,2,0 \right)&=& \frac{\sgn \left( \left \langle 10 \right \rangle  \right) \sgn \left( \left \langle 20 \right \rangle \right)}{ \left|\langle 12\rangle\right|^{2\Delta-h}\left|\langle 10\rangle\right|^{h}\left|\langle 02\rangle\right|^{h}}.
    \label{3pt_explicit}
\end{eqnarray}
Here $f^S_h$, $f^A_h$ are respectively symmetric and antisymmetric under the exchange $\left( \tau_1,\theta_1,\bar{\theta}_1 \right)\leftrightarrow \left( \tau_2,\theta_2,\bar{\theta}_2 \right)$.

Dividing the four-point function (\ref{4pt_shadow}) over the appropriate propagators to make it conformally invariant, we find:

\begin{equation}
    \mathcal F = \sum_h \int d\tau_0 d^2\theta_0 \frac{\left( A+S \sgn \tau_{12} \sgn \tau_{10} \sgn \tau_{20} \right)\left( A'+S'\sgn \tau_{34} \sgn \tau_{30}\sgn\tau_{40} \right)}{\left|\langle 12\rangle\right|^{-h}\left|\langle 10\rangle\right|^{h}\left|\langle 02\rangle\right|^{h}\left|\langle 34\rangle\right|^{h}\left|\langle 30\rangle\right|^{-h}\left|\langle 04\rangle\right|^{-h}}+O(\varepsilon^2).
    \label{W_shadow}
\end{equation}
where we denote $\mathcal W=1+\epsilon \mathcal F$.
We call the functions in the sum (\ref{W_shadow}) $\Xi_h$.
They are eigenfunctions of the Casimir:
\begin{equation}
    \mathcal C \Xi_h=h^2 \Xi_h.
    \label{C_xi}
\end{equation}
The shadow representation allows us to find the explicit form of $\Xi_h$ as an integral.
In the coordinates chosen as in (\ref{theta_fix}), (\ref{tau_fix}), the eigenfunction reads:

\begin{equation}
   \Xi_h = \int d\tau_0 d^2\theta_0 \frac{\left( A-S \sgn \chi \sgn \tau_0 \sgn \left( \chi-\tau_0 \right) \right)\left( A'-S'\sgn \left( 1-\tau_0 \right) \right)}{\left|\chi\right|^{-h}\left|\tau_0-\theta_0\bar{\theta}_0\right|^{h}\left|\chi-\tau_0-\theta_0\bar{\theta}_0\right|^{h}\left| 1-\tau_0-\theta_0\bar{\theta}_0\right|^{-h}}.
    \label{F_shadow_fix}
\end{equation}

Now we integrate over Grassmann coordinates and rename $y=\tau_0$, to find the four-point function as an integral over even coordinates:


\begin{equation}
    \Xi_h=\int dy  {\left( A-S \sgn \chi \sgn y \sgn \left( \chi-y \right) \right)\left( A'-S'\sgn \left( 1-y \right) \right)}\frac{h |\chi|^h |1-y|^h}{|y|^h |\chi-y|^h}\left( \frac 1{y} + \frac{1}{\chi-y} -\frac{1}{1-y}\right).
    \label{F_y}
\end{equation}
We break this integral into four parts in a straightforward way:

\begin{equation}
    \Xi_h = AA' \Xi_h^{AA} + AS' \Xi_h^{AS} + SA'\Xi_h^{SA}+ SS' \Xi_h^{SS}.
    \label{F_sum}
\end{equation}
Each of the four integrals can be found directly, but we can save the effort if we notice similarities to the non-supersymmetric SYK model with complex fermions.
In that case, the four-point function is given by an integral:

\begin{equation}
    \Psi_h^{\mathcal N=0}=\int dy  {\left( a+s \sgn \chi \sgn y \sgn \left( \chi-y \right) \right)\left( a'+s'\sgn \left( 1-y \right) \right)}\frac{|\chi|^h |1-y|^{h-1}}{|y|^h |\chi-y|^h}.
    \label{F_N=0}
\end{equation}

It is also a sum of four parts:

\begin{equation}
    \Psi_h^{\mathcal N=0}=aa' \Psi_h^A\left( \chi \right)+ ss' \Psi_h^S\left( \chi \right) + as' \Psi_h^{AS}\left( \chi \right)+sa' \Psi_h^{SA}\left( \chi \right).
    \label{F_N=0_sum}
\end{equation}

These functions have different parity under exchanges of two fermions. 
The function $\Psi^A$ is odd under both $1 \leftrightarrow 2$ and $3 \leftrightarrow 4$, and it is the same as the eigenfunction in the original SYK model, found in \cite{Maldacena:2016hyu}.
The function $\Psi^S$ is even under both of these permutations.
The functions $\Psi^{AS}$, $\Psi^{SA}$ have mixed parity. 
They break the time-reversal symmetry $\mathcal T$, whereas $\Psi^A$ and $\Psi^S$ preserve it.

Upon inspection, we see that the $\mathcal N=2$ eigenfunctions are linear combinations of the non-supersymmetric ones, in particular:

\begin{eqnarray}
    \Xi^{AA}_h&=& h \left( \Psi^{SA}_{h+1}\left( \chi \right) -\Psi^{AS}_h\left( \chi \right)\right),\\
    \Xi^{SS}_h&=& h \left( \Psi^{AS}_{h+1}\left( \chi \right)-\Psi^{SA}_h\left( \chi \right) \right),\\
    \label{xi_AS_n0}
    \Xi^{AS}_h&=& h \left( -\Psi^{S}_{h+1}\left( \chi \right)+\Psi^{A}_h\left( \chi \right) \right),\\
    \Xi^{SA}_h&=& h \left(- \Psi^{A}_{h+1}\left( \chi \right)+\Psi^{S}_h\left( \chi \right) \right).
    \label{N2_via_N0}
\end{eqnarray}
We notice that an eigenfunction in the $\mathcal N=2$ model built from three-point functions of the same type ($AA$ or $SS$) is a sum of ``mixed'' eigenfunctions in $\mathcal N=0$, and vice versa: a ``mixed'' $\mathcal N=2$ eigenfunction is a combination of ``pure'' $\mathcal N=0$ eigenfunctions.
As a consequence, ``mixed'' eigenfunctions in $\mathcal N=2$ preserve time-reversal, and ``pure'' four-point functions break it.
This happens because the $\mathcal N=2$ eigenfunctions are integrals over Grassmann coordinates.
The Grassmann measure $d\theta_0d\bar{\theta}_0$ is an imaginary quantity and therefore is odd under time-reversal.
So the functions of mixed parity, which are $\mathcal T$-odd in the $\mathcal N=0$ model, turn out to be $\mathcal T$-even in the $\mathcal N=2$ model.

It is interesting to notice the properties of these eigenfunctions under the transformation $h \leftrightarrow -h$. 
From (\ref{4pt_shadow}), we see that this transformation corresponds to exchange of pairs of fermions: $(1,2) \leftrightarrow (3,4)$. 
We know what happens to the eigenfunctions of the non-supersymmetric SYK when we take $h \leftrightarrow 1-h$:
\begin{eqnarray}
    \Psi^A_{1-h}&=& \Psi^A_{h},\\
    \Psi^S_{1-h}&=& \Psi^S_h,\\
    \Psi^{AS}_{1-h}&=& \Psi^{SA}_h.
    \label{psi_n=0_sym}
\end{eqnarray}
From here, we can see that:
\begin{eqnarray}
    \Xi_{-h}^{AA}&=&  \Xi_h^{AA},\\
    \Xi_{-h}^{SS}&=& \Xi_h^{SS},\\
    \Xi_{-h}^{AS}&=& \Xi_{h}^{SA}.
    \label{xi_sym}
\end{eqnarray}

The transformation exchanges the $\mathcal T$-even functions and leaves $\mathcal T$-odd functions invariant.

Since the SYK model is $\mathcal T$-invariant, in what follows we are interested in the $\mathcal T$-invariant eigenfunctions, $\Xi_h^{AS}$ and $\Xi_h^{SA}$.
Moreover, because of the relation (\ref{xi_sym})  we can focus our attention on the $\Xi^{AS}$ function only.
For brevity, we call it $\xi_h$:
\begin{equation}
   \xi_h\left( \chi \right) \equiv \Xi^{AS}_h\left( \chi \right) = h\left( \Psi^A_h\left( \chi \right)-\Psi^{S}_{h+1}\left( \chi \right) \right)=h\left( \Psi^A_h\left( \chi \right)-\Psi^S_{-h}\left( \chi \right) \right).
    \label{Xi_def}
\end{equation}

For $\chi<1$ we can express the eigenfunctions in terms of $\varphi_h$ defined in (\ref{phi_def}).

\begin{equation}
    \xi_h=
        h\left( 1+\frac{1}{\cos \pi h} \right)\varphi_h\left( \chi \right)+h\left( 1-\frac{1}{\cos \pi h}  \right)\varphi_{-h}\left( \chi \right), \qquad \chi<1.
    \label{AS_phi}
\end{equation}

For $\chi>1$, we have to do an analytical continuation.
Using the results from the $\mathcal N=0$ SYK, we find:
\begin{equation}
    \xi_h=\frac{4}{\sqrt{\pi}}\Gamma\left( 1+\frac{h}{2} \right)\Gamma\left( \frac{1-h}{2} \right)\left( \,_2F_1\left( \frac{h}{2}, \frac{1-h}{2}; \frac12; \left( \frac{2-\chi}{\chi} \right)^2 \right)+h\frac{2-\chi}{\chi} \,_2F_1\left( \frac{h}{2},\frac{1-h}{2}; \frac32; \left( \frac{2-\chi}{\chi} \right)^2 \right) \right).
    \label{AS_chi>1}
\end{equation}


We can expand a supersymmetric conformal four-point function in terms of the $\xi_h$ functions.
Schematically, the SYK four-point function looks as:
\begin{equation}
   \mathcal F=\frac{\mathcal F_0}{1-K}.
   \label{F_sum_def}
\end{equation}
The SYK kernel $K$ commutes with the $\mathcal N=2$ Casimir and therefore is diagonalized by its eigenfunctions $\xi_h$.
As our next step, we find the eigenvalues of the kernel.

\subsection{Kernel}
\label{sec:kernel_1d}
\begin{figure}
    \centering
    \includegraphics[width=.5\textwidth]{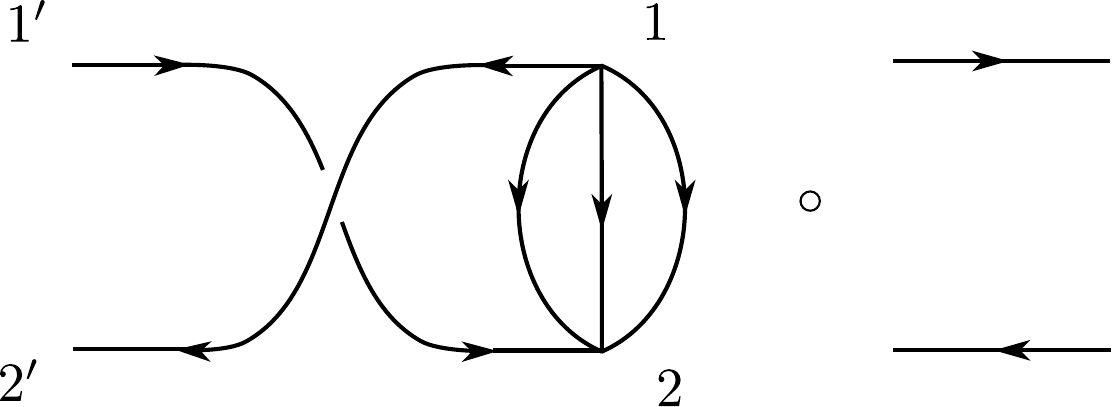}
    \caption{$\mathcal N=2$ conformal kernel.}
    \label{fig:kernel}
\end{figure}

Schematically, the $\mathcal N=2$ SYK kernel looks like fig. \ref{fig:kernel}.
Unlike the non-supersymmetric case, here chirality restricts us to only one form of the kernel operator.
The kernel in the integral form is as follows:
\begin{equation}
    K= \left( \hat{q}-1 \right)b^{\hat{q}} J \frac{\sgn  \tau_{12} }{|\langle 12 \rangle|^{2\Delta \left( \hat{q}-2 \right)}}\frac{\sgn \tau_{1'2}}{|\langle 1'2 \rangle|^{2\Delta}}\frac{\sgn \tau_{12'}}{|\langle 12' \rangle|^{2\Delta}}d\tau_1 d\tau_2 d\bar{\theta}_1 d\theta_2 . 
    \label{K_int}
\end{equation}
The kernel can act either on the 12 or on the 34 channel of the four-point function.
In the shadow representation, we construct the four-point point function as an integral of $12y$ and $34y$ three-point function, where $y$ is the arbitrary variable we integrate over.
This means that to find out how the kernel acts on a four-point function, it suffices to consider how it acts on the three-point functions.
We have fixed the form of the possible three-point functions in (\ref{fA_def}, \ref{fS_def}).
These $f_h^A$, $f_h^S$ functions diagonalize the kernel:
\begin{equation}
   \int K\left( 1',2'|1,2 \right) f^A_h\left( 1,2,0 \right)=k^A(h)f^A_h\left( 1',2',0 \right), \qquad \int K\left( 1',2'|1,2 \right) f^S_h\left( 1,2,0 \right)=k^S(h)f^S_h\left( 1',2',0 \right).
   \label{Kf=kf}
\end{equation}
To find the eigenvalues $k^A$ and $k^S$ conveniently, we first take $\tau_0$ in the three-point function to infinity, and set:
\begin{eqnarray}
    1' \to (1,\theta),\\
    2' \to \left( 0,\bar{\theta} \right).
    \label{12_fix}
\end{eqnarray}
Then the eigenvalues are given by the integrals,
\begin{equation}
    k^A= \frac{\tan \pi \Delta}{4\pi} \int d\tau_1 d\tau_2 d\bar{\theta}_1 d\theta_2 \frac{1 }{|\langle 12 \rangle|^{1-2\Delta - h}}\frac{\sgn \left( 1-\tau_2 \right)}{|\langle 1'2 \rangle|^{2\Delta}}\frac{\sgn \left( \tau_1 \right)}{|\langle 12' \rangle|^{2\Delta}}. 
    \label{kA_int}
\end{equation}
\begin{equation}
    k^S= \frac{\tan \pi \Delta}{4\pi} \int d\tau_1 d\tau_2 d\bar{\theta}_1 d\theta_2 \frac{\sgn \tau_{12} }{|\langle 12 \rangle|^{1-2\Delta - h}}\frac{\sgn \left( 1-\tau_2 \right)}{|\langle 1'2 \rangle|^{2\Delta}}\frac{\sgn \left( \tau_1 \right)}{|\langle 12' \rangle|^{2\Delta}}. 
    \label{kS_int}
\end{equation}
These integrals are of the same type we have encountered in the $\mathcal N=0$ SYK kernel.
We can make a change of variables and transform them into products of one-dimensional integrals.
The details of the computation can be found in Appendix \ref{sec:app_kernel}.
Explicitly, the answer reads:
\begin{eqnarray}
    \label{kA_ans}
    k^A&=&-\frac{1}{\pi^2} \Gamma\left( -2\Delta \right) \Gamma\left( 2-2\Delta \right) \Gamma\left( 2\Delta-h \right)\Gamma\left( 2\Delta+h \right)\sin 2\pi \Delta \left( \sin 2\pi \Delta-\sin \pi h \right),\\
    k^S&=&-\frac{1}{\pi^2} \Gamma\left( -2\Delta \right) \Gamma\left( 2-2\Delta \right) \Gamma\left( 2\Delta-h \right)\Gamma\left( 2\Delta+h \right)\sin 2\pi \Delta \left( \sin 2\pi \Delta+\sin \pi h \right).
    \label{k_c_ans}
\end{eqnarray}
These expressions coincide with the results of \cite{Peng:2017spg}, up to renaming $h \to h+1/2$.

\begin{figure}
    \centering
    \includegraphics[width=.6\textwidth]{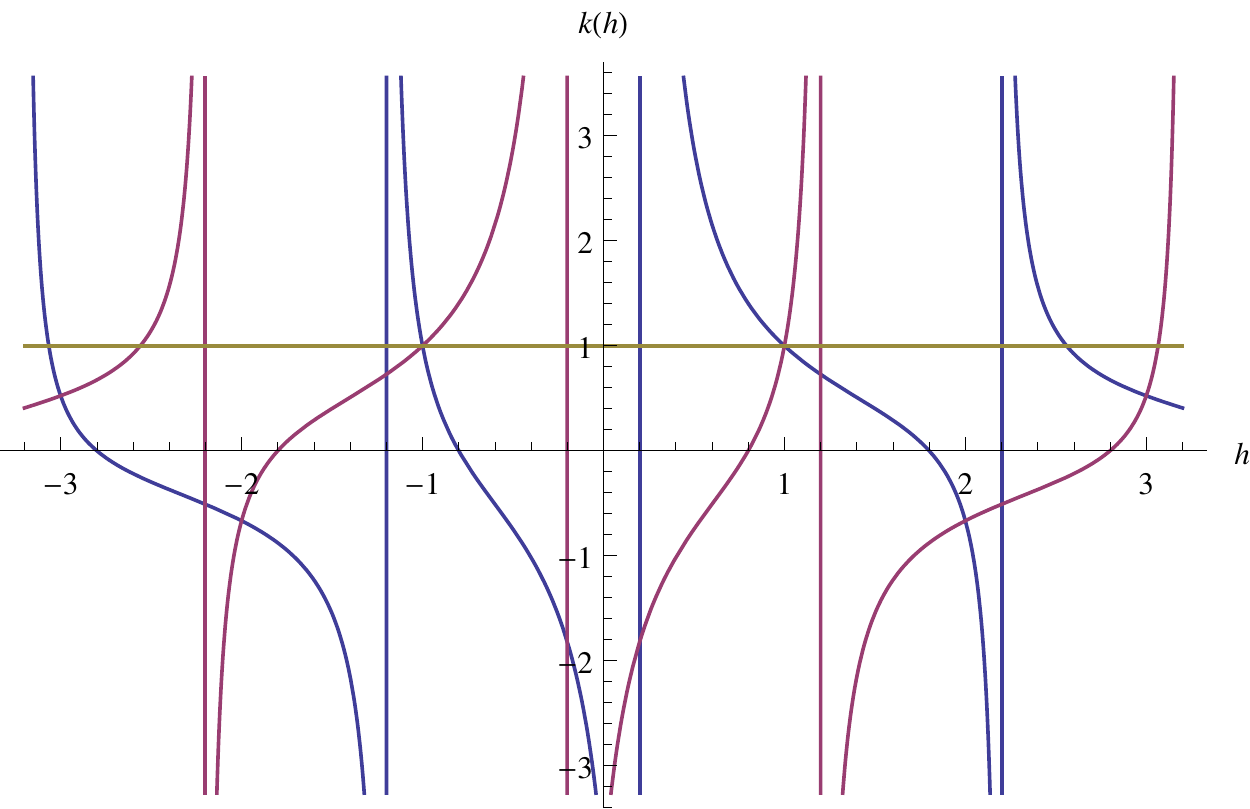}
    \caption{Eigenvalues of the antisymmetric (red) and symmetric (blue) kernels at $\hat{q}=5$.}
    \label{fig:kernel_eigen}
\end{figure}

We see that the eigenvalues satisfy:
\begin{equation}
    k^A\left( h \right)=k^S\left( -h \right).
    \label{k_h_-h}
\end{equation}
This allows for ``mixed'' four-point functions, i.e. those built from three-point functions with opposite symmetries.
The $\Xi^{AS}$ eigenfunctions, which we are going to use to expand the full four-point function, are constructed from three-point function of different types.
Acting with the kernel on the $\Xi^{AS}$ eigenfunction from the left (in the 12 channel), we multiply it by the $k^A$ eigenvalue; acting from the right, we multiply it by the $k^S$ eigenvalue. 
But if we exchange $h \leftrightarrow -h$ (transforming $\Xi^{AS}$ to $\Xi^{SA}$), we exchange the two sides in the shadow representation, and therefore exchange two channels. 
The condition (\ref{k_h_-h}) is needed to allow this transformation.

For consistency, in what follows the kernel always acts on the four-point function from the left, so that the $k^A$ eigenvalue corresponds to the $\Xi^{AS}$ eigenfunction.

The eigenvalues of the $\mathcal N=2$ kernel look very much like the eigenvalues of the non-supersymmetric kernel which we list in Appendix \ref{sec:app_n0}.
The exact relation is:
\begin{eqnarray}
    k^{A}_{\mathcal N=2}\left( h \right)=\frac{2\Delta +h-1}{2\Delta-2} k^{A}_{\mathcal N=0}\left( h \right), \\
    k^{A}_{\mathcal N=2}\left( h \right)=\frac{2\Delta-h-1}{2\Delta}k^{S}_{\mathcal N=0}\left( -h \right).
    \label{k_n2_n0}
\end{eqnarray}
The symmetry (\ref{k_h_-h}) is a direct consequence of the symmetry $h \leftrightarrow 1-h$ for the eigenvalues of non-supersymmetric kernel.

The dimensions of the operators in the theory are given by the solutions to the equation $k=1$ (see fig. \ref{fig:kernel_eigen}).
Generally these dimensions are irrational, given by an asymptotic formula:

\begin{eqnarray}
    h^A&=& 2n+1+2\Delta+O\left( \frac{1}{n} \right),\\
    h^S&=& 2n+2\Delta+O\left( \frac{1}{n} \right), \qquad n>0.
    \label{h_asymp}
\end{eqnarray}

There is also a mode with $h=1$ in both channels (which is the same as the $h=3/2$ mode of \cite{Peng:2017spg}.
This mode represents the charge multiplet, consisting of the $R$-charge, the supercharge and the stress tensor:

\begin{equation}
   \mathcal Q=R+\theta \bar{Q}+\bar{\theta} Q+\theta \bar{\theta} T. 
   \label{Q_def}
\end{equation}
Since the dimension of $\mathcal Q$ is one, the dimension of the $R$--charge operator is also one, and the dimension of the stress tensor is two, just as in the non-supersymmetric complex SYK model \cite{Bulycheva:2017uqj}.
Notice also that like the $U(1)$ charge in the non-supersymmetric model, the $R$--charge, despite being conserved, has non-zero dimension in the infrared limit.

\subsection{Inner product}
\label{sec:norm_1d}

To apply the formula (\ref{F_sum_def}) for the four-point function, we need to project the zero-rung function $\mathcal F_0$ to the basis of the Casimir eigenfunctions. 
To this end, we first find an inner product for the $\xi_h$ eigenfunctions.

For the non-supersymmetric SYK model, the eigenstates of the Casimir form a Hilbert space \cite{Maldacena:2016hyu}.
In the supersymmetric case, we should not expect this, since the eigenstates are functions of a superspace and therefore the set of states may contain functions of odd variables.
Indeed, it has been found in \cite{Murugan:2017eto}, that the $\mathcal N=1$ eigenfunctions do not form a Hilbert space.
Nevertheless, we want to get as close to a Hilbert space as possible.

An invariant inner product of chiral-antichiral four-point functions looks as follows:

\begin{equation}
   \langle f,g\rangle=\int \frac{dt_1 dt_2 d\bar{\theta}_1 d\theta_2}{\langle 12 \rangle}\frac{dt_3 dt_4 d\bar{\theta}_3 d\theta_4}{\langle 34 \rangle} f \cdot g \equiv \int d\mu\left( 1,2 \right) d\mu \left( 3,4 \right) f\cdot g.
    \label{fg_norm}
\end{equation}
Here we have defined the two-particle integration measure $d\mu\left( i,j \right)$, which is conformally invariant but not real: $d\mu\left( i,j \right)\neq d\bar{\mu}\left( i,j \right)$.
Therefore we do not expect the inner product to be real, and this is why we have $f\cdot g$ instead of $\bar{f} \cdot g$ in the inner product.
For the same reason, we do not expect the Casimir to be Hermitean with respect to this inner product.
Instead, we require it to be bilinear symmetric.

We have shown in the beginning of Section \ref{sec:4pt} that we can fix the coordinates in the four-point function, so that it does not depend on odd coordinates in the superspace.
In the same way, we can use the supergroup to make the measure a function of $\chi$ only.
The details of this calculation can be found in Appendix \ref{sec:app_norm1d}, the result being:

\begin{equation}
    \langle f,g \rangle=\int_{-\infty}^\infty \frac{d\chi}{\chi \left(1-\chi \right)} fg.
    \label{fg_norm_red}
\end{equation}
This inner product is clearly not positive-definite, so the $\mathcal N=2$ eigenstates do not form a Hilbert space.
It is easy to see that the Casimir \ref{C_def} is symmetric with respect to this norm:

\begin{equation}
    \langle \mathcal Cf, g \rangle = \int_{-\infty}^\infty d\chi f \partial_\chi \left( \chi \partial_\chi g \right)=\left. \left( f\chi\partial_\chi g- g \chi\partial_\chi f \right)\right|_{-\infty}^\infty+\int_{-\infty}^\infty f \partial_\chi\left( \chi \partial_\chi g \right)=\langle f, \mathcal C g \rangle, 
    \label{C_herm}
\end{equation}
provided that a certain boundary condition at infinity is satisfied:
\begin{equation}
    \left. \left( f\chi\partial_\chi g- g \chi\partial_\chi f \right)\right|_{-\infty}^\infty=0.
    \label{fg_inf}
\end{equation}

If the inner product (\ref{fg_norm_red}) were positive definite, we would find a complete set of functions by requiring that the eigenvalue of the Casimir $h^2$ be positive and then looking for normalizable (or continuum-normalizable) states.
We are not in this situation here.
Nevertheless we can find a set of functions with non-negative norm.
If we require that the Casimir does not bring us out of this set,
\begin{equation}
   \left \langle \xi_h, \xi_h \right \rangle \ge 0, \qquad \left \langle \mathcal C \xi_h, \mathcal C \xi_h \right \rangle \ge 0 \qquad \Rightarrow \qquad h^4 \ge 0.
   \label{xi-h_positive}
\end{equation}
then it implies that the eigenvalue of the Casimir has to be real:
\begin{equation}
   h^2 \in \mathbb R.
   \label{h2_real}
\end{equation}
In what follows, we see that the condition (\ref{h2_real}) is enough to guarantee that the inner product in the $\xi_h$ basis is positive-(semi)definite.

The eigenvalue of the Casimir can be real if $h$ is either purely imaginary or purely real. 
In the latter case, the eigenstate is normalizable only if we further restrict to integer $h$:
\begin{equation}
   h \in i \mathbb R \qquad \text{or} \qquad h \in \mathbb Z.
   \label{h_cond}
\end{equation}
The first case gives us a continuous series of states, and we expect them to be continuum-normalizable, that is their inner product is proportional to a delta function:
\begin{equation}
    \langle \xi_{is}, \xi_{is'}\rangle\sim \delta\left( s-s' \right).
    \label{<>_delta}
\end{equation}
This singular contribution comes from the vicinity of $\chi=0$:
\begin{equation}
    \langle \xi_{is}, \xi_{is'} \rangle \sim \int_{-\epsilon}^\epsilon \frac{d\chi}{\chi} \xi_{is}\xi_{is'}.
    \label{xi_xi_0}
\end{equation}
For small positive $\chi$, the Casimir eigenfunctions have a power-like behavior:
\begin{equation}
    \chi \to +0: \qquad \varphi_{is} \sim \chi^{is} B\left( is,is \right).
    \label{phi_zero}
\end{equation}

To find the asymptotic of the eigenfunction for negative $\chi$, we once again represent $\xi_h$ via  $\mathcal N=0$ eigenfunctions:
\begin{equation}
    \xi_h=h\left( \Psi^A_h-\Psi^S_{-h} \right).
    \label{xi_Psi_Psi_norm}
\end{equation}
The function $\Psi^A_h$ is symmetric under $\chi \to \frac{\chi}{\chi-1}$, and $\Psi^S_h$ is antisymmetric under the same transformation.
It means in particular that $\Psi^A_h$ is an even function of $\chi$ in the vicinity of zero, and $\Psi^S_h$ is odd.
Since the measure $d\chi/\chi$ is odd, only the terms odd in $\chi$ in the integrand of (\ref{xi_xi_0}) contribute to the final answer.
So in terms of the $\mathcal N=0$ eigenfunctions, the inner product is:
\begin{equation}
    \langle \xi_{is}, \xi_{is'} \rangle = \int_{-\epsilon}^\epsilon \frac{d\chi}{\chi} is \cdot is'\left( -\Psi^A_{is} \Psi^S_{is'+1}-\Psi^S_{is+1}\Psi^A_{is'} \right)= 2ss'\int_{0}^\epsilon \frac{d\chi}{\chi} \left( \Psi^A_{is} \Psi^S_{is'+1}+\Psi^S_{is+1}\Psi^A_{is'} \right).
    \label{<>_Psi}
\end{equation}
For small positive $\chi$, the $\Psi^A_h, \Psi^S_h$ eigenfunctions behave as follows:
\begin{eqnarray}
    \Psi^A_h &\sim& \left( 1+\frac{1}{\cos \pi h} \right)B\left( h,h \right)\chi^h + \left( 1-\frac{1}{\cos \pi h} \right)B\left( 1-h,1-h \right)\chi^{1-h},\\
    \Psi^S_h &\sim& \left( 1-\frac{1}{\cos \pi h} \right)B\left( h,h \right)\chi^h + \left( 1+\frac{1}{\cos \pi h} \right)B\left( 1-h,1-h \right)\chi^{1-h}, \qquad \chi \to +0.
    \label{Psi_chi0}
\end{eqnarray}
Bringing (\ref{<>_Psi}, \ref{Psi_chi0}) together, using the integral form of the delta-function:
\begin{equation}
    \int_{0}^\epsilon\frac{d\chi}{\chi}\left( \chi^{i\left( s-s' \right)}+\chi^{-i\left( s-s' \right)} \right)=2\pi \delta\left( s-s' \right),
    \label{delta_int}
\end{equation}
and an identity for the Euler's beta function:
\begin{equation}
    B\left( is, is \right) B\left( -is,-is \right)=\frac{4\pi}{s}\coth \pi s,
    \label{B+B-}
\end{equation}
we can find the norm for the continuous series as: 
\begin{equation}
    \langle \xi_{is}, \xi_{is'}\rangle =4\pi s \tanh \pi s \cdot 2\pi \delta\left( s-s' \right).
    \label{xi_cont_norm}
\end{equation}
In particular, this norm is real and positive for real non-zero $s$, as expected. 

The reader may be puzzled that the inner product of the basis states $\xi_{is}$ is positive definite, given that the inner product (\ref{fg_norm_red}) is not.
Indeed, we can easily find a function which has a negative norm, for example one that is close to zero for positive $\chi$ and has a bump at negative $\chi$.
How can it be expanded in the $\xi_{is}$ basis?

The matter becomes clear if we recall that the $\xi_{is}$ functions are generally complex, as are the expansion coefficients, therefore the condition that the norm be non-negative is not very restrictive.
To see this, we can break the eigenfunction into a real and an imaginary parts,
\begin{equation}
   \xi_{is} = \zeta_{s} + i\eta_{s}.
   \label{xi_complex}
\end{equation}
Its complex conjugate is also in the spectrum and has the same eigenvalue:
\begin{equation}
   \bar{\xi}_{is}=\xi_{-is}= \zeta_{s} - i\eta_{s}.
   \label{xi_conj}
\end{equation}
From the inner products for $\xi_h$,
\begin{equation}
   \langle \xi_{is}, \xi_{is'} \rangle = 4\pi s \tanh \pi s \cdot 2\pi \delta\left( s-s' \right), \qquad \langle \xi_{is}, \bar{\xi}_{is} \rangle = \langle \xi_{is}, \xi_{-is}\rangle=0,
   \label{xi_xi_bar}
\end{equation}
we can find the inner products for the real and imaginary parts separately:
\begin{equation}
   \langle \zeta_s, \zeta_{s'} \rangle = -\langle \eta_s, \eta_{s'} \rangle = 2\pi s \tanh \pi s  \cdot 2\pi \delta\left( s-s' \right), \qquad \langle \zeta_s, \eta_{s'} \rangle = 0.
   \label{zeta_eta}
\end{equation}
Hence for each eigenvalue we have two real functions $\zeta_s$ and $\eta_s$, with positive and negative norm, which are orthogonal to each other.
A function that can be expanded in the $\left( \zeta_s, \eta_s \right)$ basis, clearly can be expanded in the $\xi_{is}$ basis too, possibly with complex coefficients.

Next we find the inner product of bound states, labeled by integer eigenvalues:
\begin{equation}
    h \in \mathbb Z.
    \label{h_Z}
\end{equation}
For a state to be normalizable, we have to further restrict $h$.
For a negative integer $h$, the eigenfunction $\varphi_h$ diverges, so we have to make sure that the coefficient in front of it vanishes.
In other words, the $\xi_h=\Xi_h^{AS}$ eigenfunction is normalizable at even positive or odd negative $h$:
\begin{equation}
   h^{AS} \in 2\mathbb Z_+ \qquad \text{or} \qquad h^{AS} \in 2 \mathbb Z_-+1.
    \label{h_discrete}
\end{equation}
But the spectrum should be symmetric under $h \leftrightarrow -h$. 
So for the $\Xi_h^{SA}$ eigenfunction, the choice is exactly opposite:
\begin{equation}
    h^{SA} \in 2\mathbb Z_++1 \qquad \text{or} \qquad h^{SA} \in 2 \mathbb Z_-.
    \label{h_discrete_SA}
\end{equation}

To find the norm of a bound state, we take the integral:
\begin{equation}
   \langle \xi_h, \xi_h \rangle =\int_{-\infty}^\infty \frac{d\chi}{\chi \left( 1-\chi \right)}\xi_h^2\left( \chi \right).
    \label{norm_int_bound}
\end{equation}
This integral is generally tricky, but we can express it via the norm for the bound state in the non-supersymmetric model (details in Appendix \ref{sec:app_norm_bound}). 
The result is:

\begin{equation}
    \langle \xi_h, \xi_{h'}\rangle=\delta_{hh'} 4\pi^2 |h|.
    \label{norm_bound}
\end{equation}
Again, we see that the norm is positive, except for the $h=0$ mode which has a zero norm.

The continuous set $\xi_{is}$ is orthogonal to the discrete series $\xi_n$ since for these two cases the eigenvalues of the Casimir are different.

If we were working in a true Hilbert space, the eigenstates of the Casimir with real eigenvalues would form a complete set.
If $\xi_{is}$ formed a complete set, then naively, given the inner products (\ref{xi_cont_norm}, \ref{norm_bound}), the following identity would hold:
%
\begin{equation}
   \int_{-\infty}^\infty \frac{ds}{2\pi}\frac{1}{4\pi s \tanh \pi s} \xi_{is}\left( \chi \right) \xi_{is}\left( \chi' \right)+\sum_{h \in \mathbb Z_+} \frac{1}{4\pi^2 h}\xi_{h}\left( \chi \right)\xi_{h}\left( \chi' \right) \overset{?}{=} \chi\left( 1-\chi \right)\delta\left( \chi-\chi' \right).
   \label{completeness_final}
\end{equation}
Then we can integrate both sides of this relation with a function we want to expand in the $\xi$ basis.

However, this expression cannot be correct.
The integral over the continuous states has a double pole at $s=0$ and therefore the left hand side diverges.
The root of the problem is that the our functions are not a complete set, because the constant function is orthogonal to all of them.
The constant function belongs to both the continuous and the discrete series and is a limit of $\xi_{is}$ at zero $s$: 
\begin{equation}
   \xi_0=\lim_{h\to 0}\xi_h=4.
   \label{xi_0_const}
\end{equation}
From (\ref{xi_cont_norm}) and (\ref{norm_bound}) we see that it is orthogonal to all the eigenstates.


%

We do not know a general completeness relation for these functions, but for our application it is sufficient to know the expansion of the zero-rung function, that is the relation (\ref{completeness_final}), convolved with $\mathcal F_0$. 
In Section \ref{sec:0rung}, we find that the relation (\ref{completeness_final}) convolved with $\mathcal F_0$ is true, provided the integration contour goes to the right of the double pole at $s=0$.

Another function which is orthogonal to our set is:
\begin{equation}
   \left.\frac{d}{ds}\xi_{is}\right|_{s=0}=4 \log \chi.
   \label{d/ds_log}
\end{equation}
We see that a constant and a logarithmic function lie outside of our basis.
As we have already mentioned before, we should not a priori expect the eigenfunctions of the Casimir to be a complete set of functions if the inner product is not positive-definite.
\subsection{Zero-rung four-point function and the \(h=0\) mode}
\label{sec:0rung}
\begin{figure}
   \centering
   \includegraphics[width=.3 \textwidth]{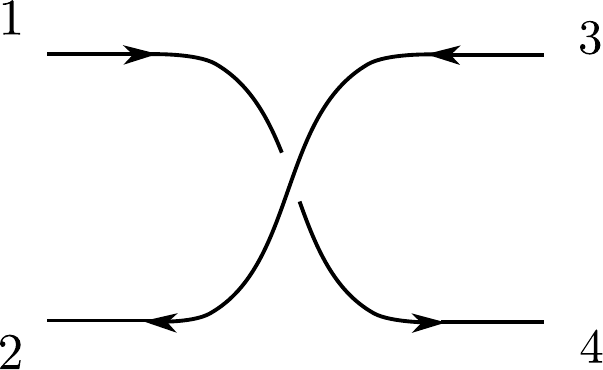}
   \caption{Zero--rung four-point function.}
   \label{fig:F0}
\end{figure}

To find the full four-point function, we project the zero-rung function $\mathcal F_0$ (see fig. \ref{fig:F0}) to the basis of the Casimir eigenfunctions $\xi_h$ using the completeness relation (\ref{completeness_final}).
Schematically, this expansion is written as:
\begin{equation}
   \mathcal F_0=\sum_h \frac{\left \langle \xi_h, \mathcal F_0 \right \rangle }{\left \langle \xi_h, \xi_h \right \rangle }\xi_h.
   \label{F0_sum_naive}
\end{equation}
The ``sum'' over $h$ includes the discrete sum over the bound states as well as the integral over the continuous series of states.
But with the latter, we run into a problem.
The integration measure in the completeness relation (\ref{completeness_final}) has a double pole at $s=0$.
To make the integral meaningful, we have to deform the integration contour away from the origin.
The result might depend on this deformation.
To see whether the procedure makes sense, we will consider the expansion of the zero-rung four-point function near $\chi=0$.

The zero-rung four-point function is a (conformally invariant) combination of conformal propagators.
Chirality restricts its form to be (see fig. \ref{fig:F0}):
\begin{equation}
    \mathcal F_0\equiv\frac{\mathcal G\left( \langle 14 \rangle \right) \mathcal G\left( \langle 32 \rangle \right)}{\mathcal G\left( \langle 12 \rangle \right) \mathcal G\left( \langle 34 \rangle \right)}=\sgn \chi \cdot |\chi|^{2\Delta}.
    \label{F0_def}
\end{equation}
The zero-rung function has a finite norm and therefore belongs to our pseudo-Hilbert space:
\begin{equation}
   \left \langle \mathcal F_0, \mathcal F_0 \right \rangle = \mathrm{p. v.}\int \frac{d\chi}{\chi\left( 1-\chi \right)}|\chi|^{4\Delta}<\infty. 
   \label{F0_norm}
\end{equation}
The inner product of an eigenfunction with the zero-rung propagator is related to the eigenvalue of the kernel, in full analogy with the non-supersymmetric case:

\begin{equation}
    \langle \xi_h, \mathcal F_0 \rangle = \frac12\alpha k^A(h), 
    \label{0rung_kernel}
\end{equation}

where $\alpha$ is similar to the $\alpha_0$ coefficient in the non-supersymmetric model:
\begin{equation}
    \frac{1}{\alpha}=b^{\hat{q}}J \left( \hat{q}-1 \right)=\frac{1-2\Delta}{8 \pi \Delta}\tan \pi \Delta.
    \label{alpha_def}
\end{equation}
The computation can be found in Appendix \ref{sec:app_0rung}.

To expand the zero-rung four-point function, we have to first determine whether it has the symmetry of $AS$ or $SA$ type.
If it has the symmetry of the $AS$ type, it expands in the $\Xi^{AS}=\xi_h$ basis:
\begin{multline}
    \mathcal F_0^{AS}\left( \chi \right)=\alpha \int_{-\infty}^\infty \frac{ds}{2\pi} \frac{1}{4\pi h \tan \pi h}{k^A\left( h \right)}\Xi^{AS}_h\left( \chi \right)+\\
    \alpha\sum_{h \in 2\mathbb Z_+}\frac{1}{4\pi^2 |h|}k^A\left( h \right)\Xi^{AS}_h\left( \chi \right)+\alpha\sum_{h \in 1-2\mathbb Z_+}\frac{1}{4\pi^2 |h|}k^A\left( h \right)\Xi^{AS}_h\left( \chi \right).
    \label{F0^AS_series}
\end{multline}
Here in the integral we take $h=is$. 
For integer $h$, we can use an identity:
\begin{equation}
   k^A(h)=k^A(-h), \qquad h\in \mathbb Z,
   \label{kA(int)}
\end{equation}
and rewrite (\ref{F0^AS_series}) as:
\begin{equation}
    \mathcal F_0^{AS}\left( \chi \right)=\alpha \int_{-\infty}^\infty \frac{ds}{2\pi} \frac{1}{4\pi h \tan \pi h}{k^A\left( h \right)}\xi_h\left( \chi \right)+
    \alpha\sum_{h \in \mathbb Z_+}\frac{1}{4\pi^2 h}k^A\left( h \right)\xi_h\left( \chi \right).
    \label{F0_1d_full}
\end{equation}
If however the zero-rung four-point function has the symmetry of the $SA$ type, it expands in terms of $\Xi^{SA}$ functions:
\begin{multline}
    \mathcal F_0^{SA}\left( \chi \right)=\alpha \int_{-\infty}^\infty \frac{ds}{2\pi} \frac{1}{4\pi h \tan \pi h}{k^S\left( h \right)}\Xi^{SA}_h\left( \chi \right)+\\
    \alpha\sum_{h \in -2\mathbb Z_+}\frac{1}{4\pi^2 |h|}k^S\left( h \right)\Xi^{SA}_h\left( \chi \right)+\alpha\sum_{h \in 2\mathbb Z_+-1}\frac{1}{4\pi^2 |h|}k^S\left( h \right)\Xi^{SA}_h\left( \chi \right).
    \label{F0^SA_series}
\end{multline}
However, using the fact that $\Xi^{AS}_h=\Xi^{SA}_{-h}$ and (\ref{kA(int)}), we can see that these two expansions give exactly the same result:
\begin{equation}
   \mathcal F_0=\mathcal F_0^{AS}=\mathcal F_0^{SA}.
   \label{F0=}
\end{equation}

\begin{figure}
   \centering
   \includegraphics[width=.7 \textwidth]{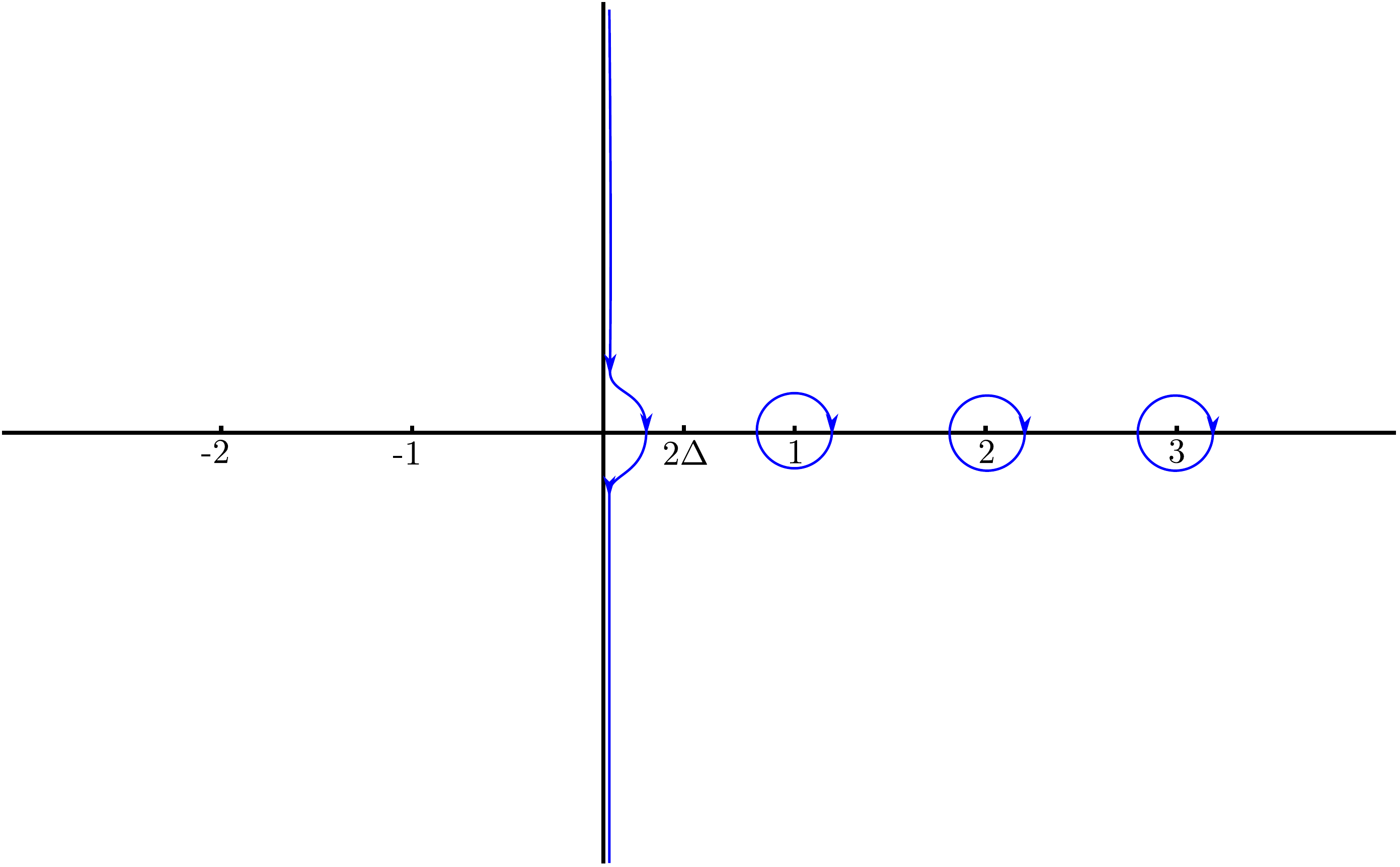}
   \caption{The integration contour for the $\mathcal N=2$ SYK model avoids the double pole at zero.}
   \label{fig:contour_2}
\end{figure}

The expression (\ref{F0_1d_full}) is a more explicit version of (\ref{F0_sum_naive}).
As we discussed before, the integration measure has a double pole  at $h=0$.
To resolve this problem, we deform the contour so that it avoids zero as in fig. \ref{fig:contour_2}.
But this deformation might add to the zero-rung four-point function a contribution of the form:

\begin{equation}
   \Res_{s=0} \frac{1}{s \tanh \pi s}\xi_{is} \sim \left. \frac{d}{ds}\xi_{is}\left( \chi \right)\right|_{s=0} \sim \log \chi.
   \label{d/ds}
\end{equation}

To see if this is the case, we look at the four-point function near $\chi=0$.
In this limit,
\begin{equation}
   \xi_h \sim h B\left( h,h \right)\left( 1+\frac{1}{\cos \pi h} \right)\chi^h + h B(-h,-h) \left( 1-\frac{1}{\cos \pi h} \right)\chi^{-h}, \qquad \chi \sim +0. 
   \label{xi_chi_0}
\end{equation}
Using the simple identity,
\begin{equation}
   k^A(h)\xi_h+k^A(-h)\xi_{-h}=\frac12 \left( k^A(h)+k^A(-h) \right)\left( \xi_h+\xi_{-h} \right)+\frac12 \left( k^A(h)-k^A(-h) \right)\left( \xi_h-\xi_{-h} \right),
   \label{k_xi_sym}
\end{equation}
we can recast (\ref{F0_1d_full}) in the form:
\begin{equation}
   \mathcal F_0=\int_C \frac{ds}{2\pi} \frac{1}{8\pi\tan \pi h}\frac{16 \Delta}{\tan \pi \Delta}B(h,h)B(2\Delta-h,2\Delta+h)B(4\Delta,-2\Delta)\left( \sin \pi h -\frac{\sin 2\pi \Delta}{\cos \pi h}\right)\chi^h+\sum_{h\in \mathbb Z_+}\left( \dots \right),
   \label{int_chi^h}
\end{equation}
where the sum in parentheses is the sum over residues of the integrand at positive integer $h$, and the contour $C$ goes as in fig.  \ref{fig:contour_2}, crossing the horizontal axis between the origin and $2\Delta$.
Closing the integration contour to the right, we find that $\mathcal F_0$ is given by a sum of residues of the integrand at the points where the kernel is singular:
\begin{equation}
   \mathcal F_0=-\Res_{h \in \mathbb Z_++2\Delta} \frac{1}{8\pi\tan \pi h}\frac{16 \Delta}{\tan \pi \Delta}\frac{\Gamma^2(h)\Gamma\left( 2\Delta-h \right)\Gamma\left( 2\Delta+h \right)\Gamma\left( -2\Delta \right)}{\Gamma\left( 2h \right)\Gamma\left( 2\Delta \right)}\left( \sin \pi h -\frac{\sin 2\pi \Delta}{\cos \pi h}\right)\chi^h.
   \label{int_res}
\end{equation}
In the leading order, this reduces exactly to the zero-rung four-point function:
\begin{equation}
   \mathcal F_0=\chi^{2\Delta}+O\left( \chi^{1+2\Delta} \right).
   \label{F0_0}
\end{equation}
If instead we had deformed the contour to lie to the left of the origin, we would have picked up a contribution proportional to $\sim \log \chi$.
We have also checked (\ref{int_res}) numerically for any $\chi$.

\begin{figure}
   \centering
   \begin{tabular}{lr}
      \includegraphics[width=.4\textwidth]{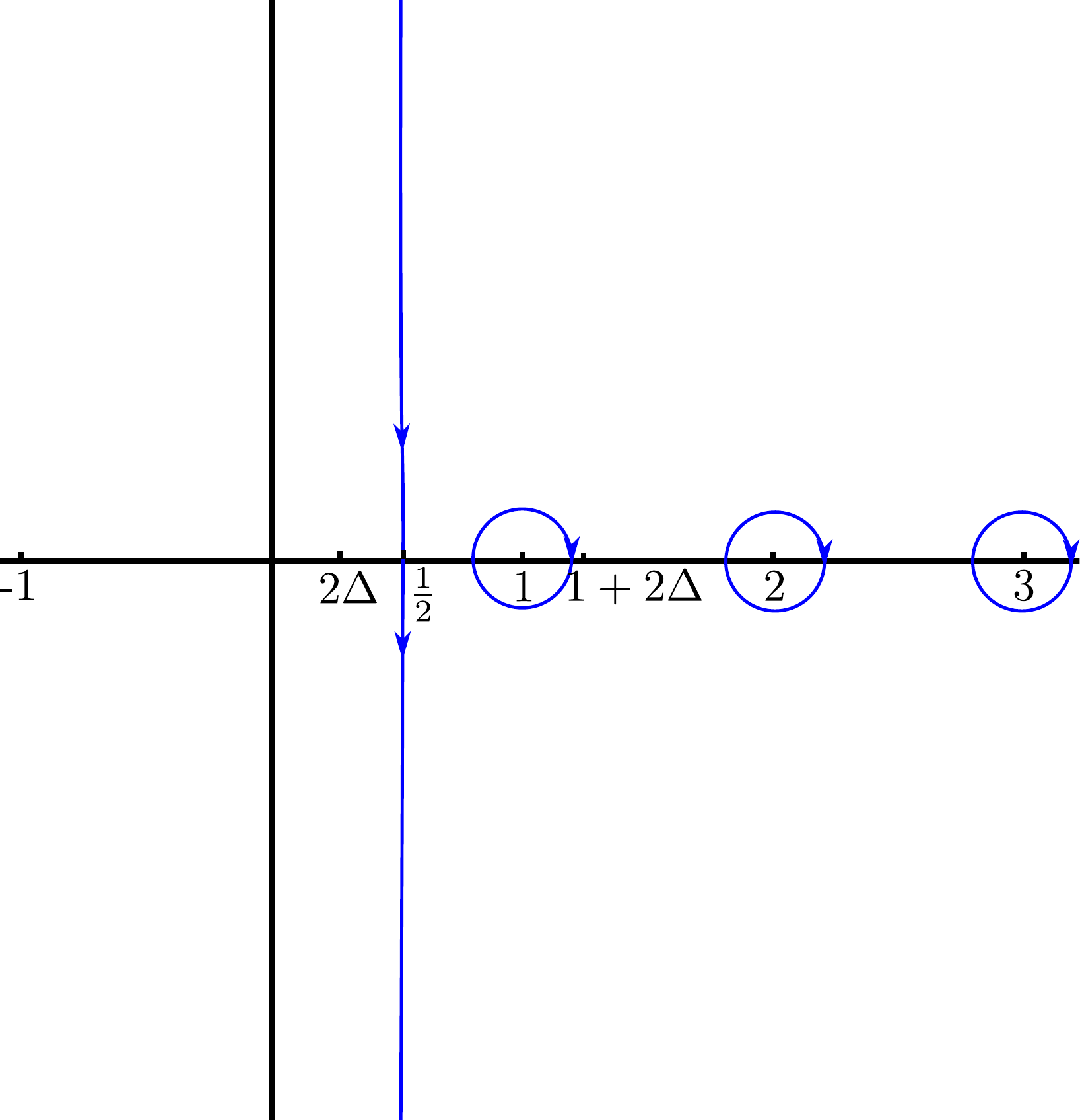}&
      \includegraphics[width=.4\textwidth]{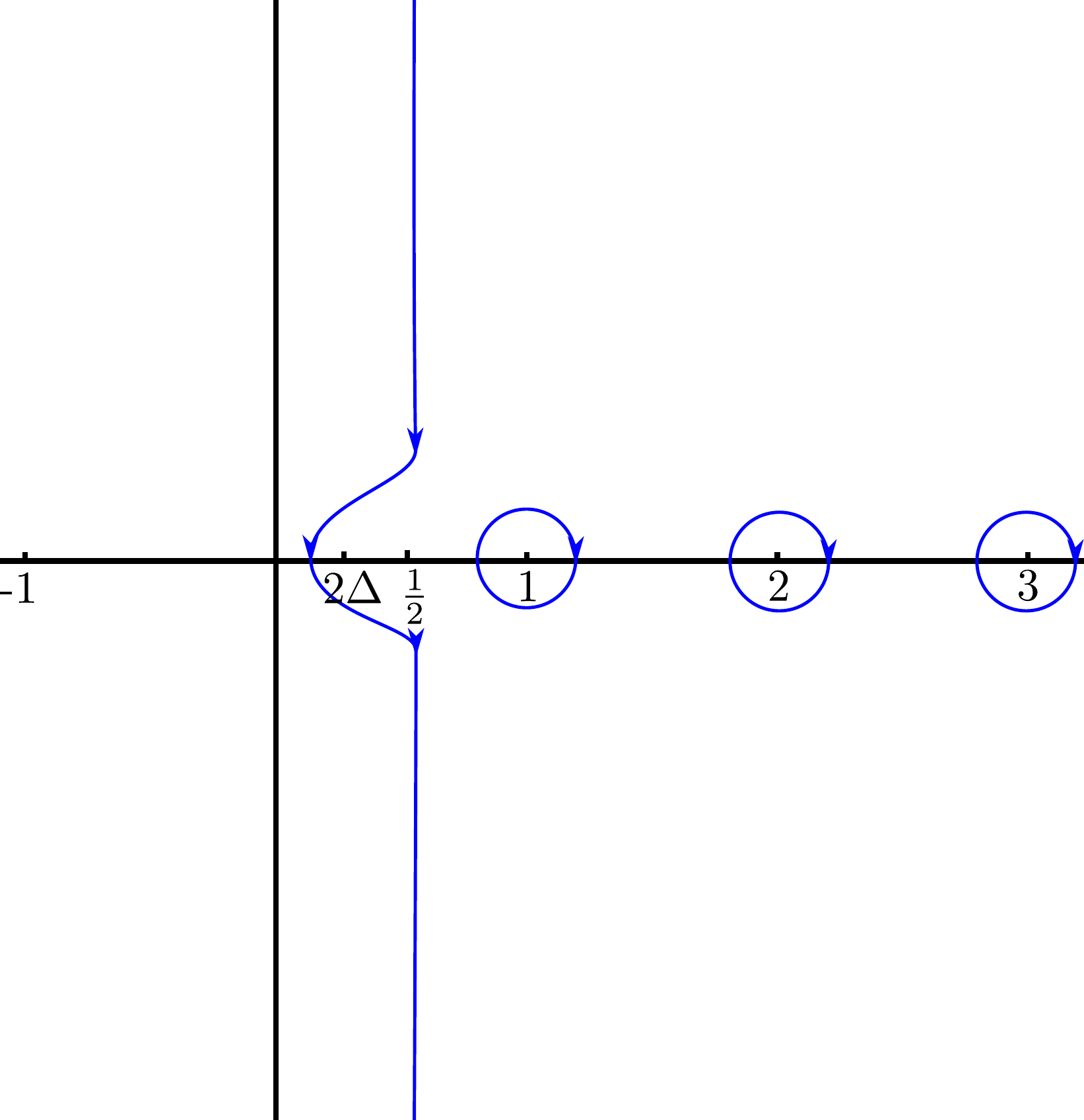}
   \end{tabular}
   \caption{Integration contours in the non-supersymmetric SYK model. The anti-symmetric channel is on the left, the symmetric one is on the right.}
   \label{fig:n0_contour}
\end{figure}

It is instructive to see how the integration contour is deformed in the non-supersymmetric SYK.
Its continuous series is at $h=\frac12+is$, so the naive integration contour is parallel to the $y$ axis and intersects the horizontal axis at $h=\frac12$.
If fermions are complex, there are two distinct channels and two distinct zero-rung four-point functions.
In the anti-symmetric channel (where the usual SYK with real fermions lives), the  zero-rung four-point function is:
\begin{equation}
   \mathcal F_0^A\left( \mathcal N=0 \right)=-\sgn\left( \chi \right)|\chi|^{2\Delta}+\sgn \chi \sgn \left( 1-\chi \right)\left|\frac{\chi}{\chi-1}\right|^{2\Delta}.
   \label{F0_A_n0}
\end{equation}
This function has a finite norm in the $\mathcal N=0$ inner product.
Near zero, this reduces to:
\begin{equation}
   \mathcal F_0^A\left( \mathcal N=0 \right)\sim -\chi^{2\Delta+1}, \qquad \chi \sim +0.
   \label{F0_A_0}
\end{equation}
Then, for the expansion in the Casimir eigenfunctions to work, we should make sure that the pole at $h=2\Delta+1$ is inside the contour.
And for the naive contour at $h=\frac12 + is$, this is automatically satisfied.

The four-point function in the symmetric channel, however,
\begin{equation}
   \mathcal F_0^S\left( \mathcal N=0 \right)=-\sgn\left( \chi \right)|\chi|^{2\Delta}-\sgn \chi \sgn \left( 1-\chi \right)\left|\frac{\chi}{\chi-1}\right|^{2\Delta},
   \label{F0_S_n0}
\end{equation}
has infinite norm and therefore does not belong to the Hilbert space.
Therefore to find a sensible expansion, we have to deform the contour.
Near zero, the symmetric zero-rung function behaves as:
\begin{equation}
   \mathcal F_0^S\left( \mathcal N=0 \right)\sim -\chi^{2\Delta}, \qquad \chi \sim +0.
   \label{F0_S_0}
\end{equation}
So to find it in the expansion, we have to make the contour go around the $h=2\Delta$ pole.
We deform it as in fig. (\ref{fig:n0_contour}), making it intersect the horizontal axis between zero and $2\Delta$.

Note that for the $\mathcal N=0$ SYK, $2\Delta$ is always smaller than $\frac12$. 
So in the symmetric channel, we need to shift the contour by a finite distance.
This reflects the fact that the symmetric zero-rung function is outside the Hilbert space.
In the $\mathcal N=2$ model, the zero-rung function belong to the pseudo-Hilbert space ``marginally'', that is the integral (\ref{F0_norm}) is convergent only in the principal value prescription.
Accordingly, the $\mathcal N=2$ integration contour also gets displaced by an infinitesimally small amount, to avoid the origin.

%
%

\subsection{General form of the four-point function}

Now we have all the ingredients needed to expand the SYK four-point function.
Formally, it is represented as:
\begin{equation}
   \mathcal F\left( \chi \right)=\sum_h \frac{\mathcal F_0}{1-K}=\sum_h \frac{1}{1-k^{A}\left( h \right)}\frac{\langle \xi_h, \mathcal F_0 \rangle}{\langle \xi_h, \xi_h \rangle} \xi_h\left( \chi \right).
    \label{F_sum_xi}
\end{equation}
Using the expansion of the zero-rung function (\ref{F0_1d_full}) allows us to write it in the form:
\begin{equation}
   \mathcal F\left( \chi \right)=-\alpha \int_C \frac{dh}{2\pi i} \frac{1}{4\pi h \tan \pi h}\frac{k^A\left( h \right)}{1-k^A\left( h \right)}\xi_h\left( \chi \right)+
    \alpha\sum_{h \in \mathbb Z_+}\frac{1}{4\pi^2 h}\frac{k^A\left( h \right)}{1-k^A\left( h \right)}\xi_h\left( \chi \right),
    \label{F_1d_full}
\end{equation}
with the integration contour $C$ being deformed as in fig. \ref{fig:contour_2} to avoid the double pole at the origin.
The integral in this expression is given by the sum of the poles in the integrand.
The poles coming from the measure are at the integer values of $h$, and are cancelled out by the sum in (\ref{F_1d_full}).
The only poles left are the ones coming from the solutions of $k(h)=1$:
\begin{equation}
   \mathcal F\left( \chi \right)=-\sum_m\Res_{h=h_m>0}\alpha \frac{1}{4\pi h \tan \pi h}\frac{1}{1-k^A\left( h \right)}\xi_h\left( \chi \right), \qquad k^A(h_m)=1.
    \label{F_1d_res}
\end{equation}
These solutions correspond to the dimensions of the physical operators in the model.
There is also an $h=1$ subspace which produces a divergence in the four-point function, since $h=1$ corresponds to the physical operator of supercharge.
This subspace should be treated separately by considering the theory outside the conformal limit.
We hope to discuss this matter elsewhere.

\section{Retarded kernel}
\label{sec:retarded}
The next question we address is the Lyapunov exponents of the modes.
To find them we introduce the retarded kernel.
We make time $\tau$ periodic with period $\beta=2\pi$ and then continue to the complex plane.
We take the left rail of the ladder diagram to be at complex time $it$ and the right rail at $\left( it+\pi \right)$, so that there is a phase difference of half a period between them.

Generally, the propagator in complex time is:
\begin{equation}
    \mathcal G_c\left( 1|2 \right)=\frac{b \sgn \left( \tau_1-\tau_2 \right)}{|\langle 12 \rangle|^{2\Delta}}\qquad \rightarrow \qquad  \mathcal G_c\left(1|2  \right)=\frac{b \left(\sgn \Re\left( \tau_1-\tau_2 \right)\right)^{2\Delta+1}}{\langle 12 \rangle^{2\Delta}}.
    \label{G_complex}
\end{equation}

\begin{figure}
    \centering
    \includegraphics[width=.4\textwidth]{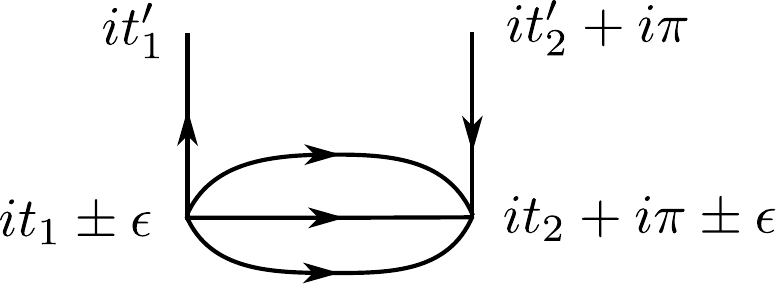}
    \caption{The retarded kernel. Retarded propagators go along rails, and the left-right propagator goes between rails.}
    \label{fig:retarded}
\end{figure}

The kernel is constructed of the propagators of two types (see fig. \ref{fig:retarded}). 
One is the conventional retarded propagator, which goes along a rail of the ladder:

\begin{equation}
   \mathcal G_R\left( 1|1' \right)=\Theta\left( t_1-t_{1'} \right)\left( \mathcal G\left( -\epsilon+it_1,it_{1'} \right)-\mathcal G\left( \epsilon+it_1,it_{1'} \right) \right)=\Theta\left( t_1-t_{1'} \right) \frac{2b \cos \pi \Delta}{\langle 11' \rangle_{ll}^{2\Delta}}.
    \label{G_R_def}
\end{equation}
Here $\langle 11' \rangle$ is the supersymmetric invariant distance between two points on the left rail of the ladder.
The other goes between the two rails of the ladder:

\begin{equation}
    \mathcal G_{lr}\left( 1|2 \right)=\frac{b}{\langle 12 \rangle_{lr}^{2\Delta}},
    \label{G_lr_def}
\end{equation}
where $\langle 12 \rangle_{lr}$ is the invariant distance between two points on the left and on the right rail.

%
%

To make time periodic, we do a conformal transformation which takes $t \to \exp(-t)$. 
Keeping in mind that the odd variables $\theta$ have conformal weight 1/2, we write the new transformed super-coordinates as follows:

\begin{equation}
    \begin{array}{llcrr}
        \tau_1= e^{-t_1},&&\qquad& \tau_2=e^{-t_2-i\pi}=-e^{-t_2},&\\
        \theta_1=e^{-\frac{t_1}{2}}\vartheta_1,&\text{(left rail)}&\qquad&\theta_2=e^{-\frac{t_2+i\pi}{2}}=-ie^{-\frac{t_2}{2}}\vartheta_2,&\text{(right rail)}\\
        \bar{\theta}_1=e^{-\frac{t_1}{2}}\bar{\vartheta}_1,&&\qquad&\bar{\theta}_2=e^{-\frac{t_2-i\pi}{2}}=ie^{-\frac{t_2}{2}}\bar{\vartheta}_2.&
    \end{array}
    \label{coord_1d}
\end{equation}

In these new coordinates, the invariant distances are as follows:
\begin{equation}
   \langle 11' \rangle_{ll}=e^{-\frac{t_1+t_1'}{2}}\left( 2\sinh \frac{t_1-t_1'}{2}-2\bar{\vartheta}_1 \vartheta_1' -\vartheta_1 \bar{\vartheta}_1-\vartheta_1' \bar{\vartheta}_1' \right),
    \label{<>_l}
\end{equation}
for the left-left invariant, and:

\begin{equation}
    \langle 12 \rangle_{lr}=\tau_1-\tau_2-2\bar{\theta}_1 \theta_2-\theta_1 \bar{\theta}_1 - \theta_2 \bar{\theta}_2=e^{-\frac{t_1+t_2}{2}}\left( 2\cosh \frac{t_1-t_2}{2}+2i\bar{\vartheta}_1\vartheta_2 -e^{-\frac{t_1-t_2}{2}}\vartheta_1 \bar{\vartheta}_1 -e^{\frac{t_1-t_2}{2}}\vartheta_2 \bar{\vartheta}_2 \right),
    \label{<>_lr}
\end{equation}
for the left-right invariant.
The reparameterization invariance of the propagator:
\begin{equation}
   \mathcal G\left( t_1,t_2 \right)=\mathcal G\left( \tau_1,\tau_2 \right)\left( \frac{d\tau_1}{dt_1}\frac{d\tau_2}{dt_2} \right)^{\Delta},
   \label{G(t)_reparam}
\end{equation}
allows us to write the retarded and the left-right propagators in the following form:
 
\begin{eqnarray}
   \label{G_R_per}
   \mathcal G_R\left( 1|1' \right) &= &\Theta\left( t_1-t_{1'} \right)\frac{2b\cos \pi \Delta}{\left( 2\sinh \frac{t_1-t_{1'}}{2}-2\bar{\vartheta}_1 \vartheta_{1'} -\vartheta_1 \bar{\vartheta}_1-\vartheta_{1'} \bar{\vartheta}_{1'} \right)^{2\Delta}},\\
   \mathcal  G_{lr}\left( 1|2 \right) &= &\frac{b}{\left( 2\cosh \frac{t_1-t_2}{2}+2i\bar{\vartheta}_1\vartheta_2 -e^{-\frac{t_1-t_2}{2}}\vartheta_1 \bar{\vartheta}_1 -e^{\frac{t_1-t_2}{2}}\vartheta_2 \bar{\vartheta}_2 \right)^{2\Delta}}.
   \label{G_lr_per}
\end{eqnarray}

%

The retarded kernel is constructed out of retarded and left-right propagators:
\begin{equation}
    K_r\left( 1',2'|1,2 \right)=\left( \hat{q}-1 \right)J \mathcal G_{R}\left( 1|1' \right) \mathcal G_R\left( 2'|2 \right) \mathcal G_{lr}^{\hat{q}-2}\left( 1|2 \right)ie^{\frac12 \left( t_1+t_2 \right)}dt_1 dt_2 d\bar{\vartheta}_1 d\vartheta_2.
    \label{Kr_def}
\end{equation}
The factor of $ie^{\frac12 \left( t_1+t_2 \right)}$ comes from the transformation of the measure.
Using the propagators (\ref{G_R_per}, \ref{G_lr_per}), we can write the kernel as follows:
\begin{equation}
   K_r\left( 1',2'|1,2 \right)=4\cos^2 \pi \Delta\left( \hat{q}-1 \right) Jb^{\hat{q}} i e^{\Delta\left( t_1+t_2 \right)}e^{-\Delta\left( t_{1'}+t_{2'} \right)}\frac{\Theta\left( t_1-t_{1'} \right)\Theta\left( t_2-t_{2'} \right)}{\langle 11' \rangle ^{2\Delta}\langle 2'2 \rangle ^{2\Delta}\langle 12 \rangle ^{1-4\Delta}}.
   \label{Kr_<>}
\end{equation}

Now we diagonalize the retarded kernel, essentially in the same way we did the conformal kernel in Section \ref{sec:kernel_1d}.
The eigenfunctions of the retarded kernel are the same three-point functions of the model (\ref{fA_def}, \ref{fS_def}).
In complex time, there is no difference between symmetric and antisymmetric eigenfunctions.
Taking the third coordinate of the three-point function to infinity, we write the kernel eigenfunction as:
\begin{equation}
    f_r^A\left( 1,2,\infty \right)=f_r^S\left( 1,2,\infty \right)=e^{-\Delta\left( t_1+t_2 \right)}\frac{1}{\langle 12\rangle^{2\Delta-h}}.
    \label{3pt_ret}
\end{equation}
Integrating over the odd variables in the expression:
\begin{equation}
    \int K_r\left( 1',2'|1,2 \right)f_r\left( 1,2,\infty \right)=k_r f_r\left( 1,2,\infty \right),
    \label{Kr_f}
\end{equation}
and fixing $\tau_1'=0$, $\tau_2'=1$,
we find that the eigenvalue is given by the integral of the same kind as for the conformal kernel:
\begin{equation}
    k_r=\left( \hat{q}-1 \right)Jb^{\hat{q}}2\left( 1-2\Delta-h \right)\left( 2\cos \pi \Delta \right)^2\int d\tau_1 d\tau_2\frac{\theta\left( -\tau_1 \right)\theta\left( \tau_2-1 \right)}{|\tau_{12}|^{2-2\Delta-h}|\tau_{1}|^{2\Delta} |\tau_{2}|^{2\Delta}}.
    \label{kr_int}
\end{equation}
Taking the integral, we find:
\begin{equation}
    k_r=\frac{\Gamma\left( -2\Delta \right)}{\Gamma\left( 2\Delta-1 \right)}\frac{\Gamma\left( -h+2\Delta \right)}{\Gamma\left( 1-h-2\Delta \right)}.
    \label{kr_ans}
\end{equation}

\begin{figure}
    \centering
    \includegraphics[width=.6\textwidth]{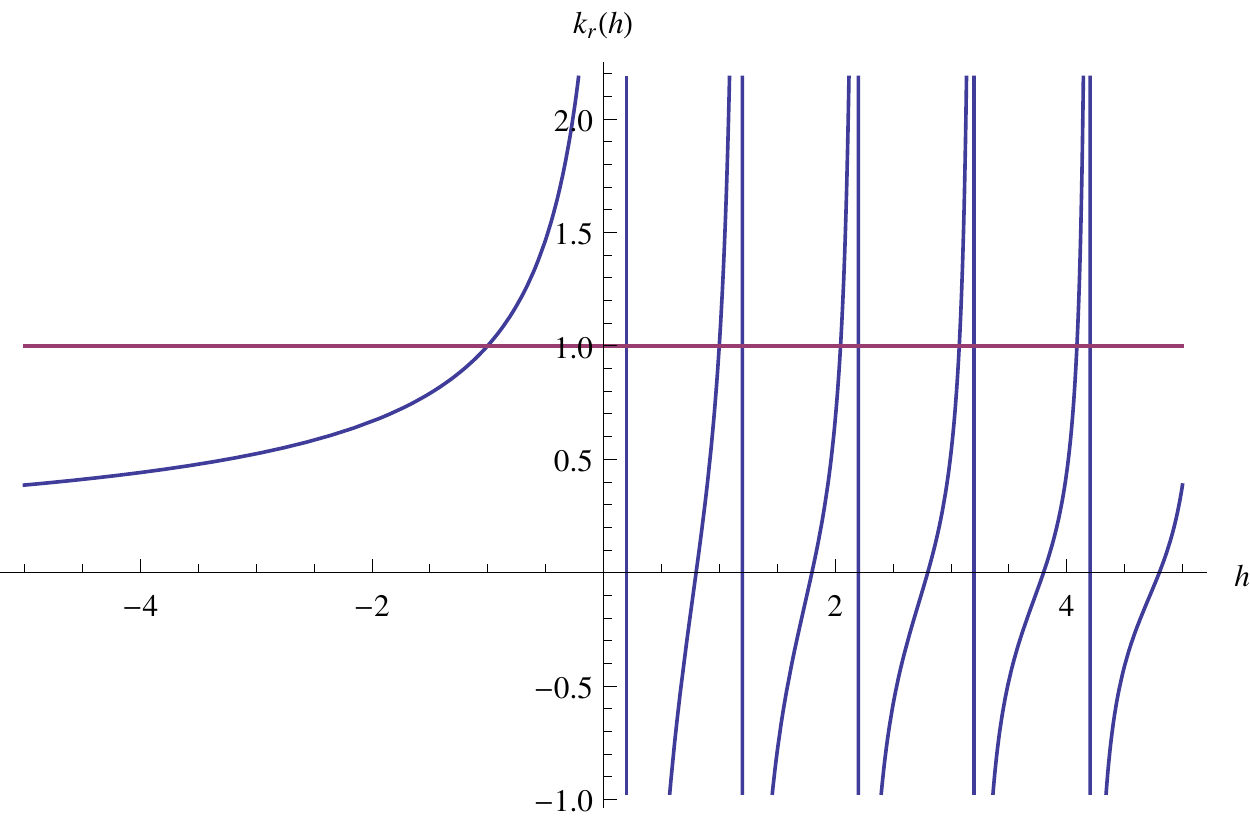}
    \caption{Eigenvalues of the retarded kernel at $\hat{q}=5$.}
    \label{fig:ret_eigen}
\end{figure}

This eigenvalue is plotted in fig. \ref{fig:ret_eigen}.
The modes potentially contributing to chaos satisfy $k_r=1$. 
The minimal weight $h$ that satisfies this constraint is $h=-1$:

\begin{equation}
    k_r|_{ h=-1 }=1 \qquad \text{for all $\Delta$}.
    \label{kr_-1}
\end{equation}

At large times, the three-point function $f_r\left( 1,2,\infty \right)$ grows (or decays) exponentially:

\begin{equation}
    f_r\left( 1,2,\infty \right)\sim e^{-ht},
    \label{f_exp}
\end{equation}
therefore the $h=-1$ mode shows maximally chaotic behavior. 
All the other modes have positive $h$ and do not contribute to the exponential growth.

\section{Generalization to two dimensions}
\label{sec:2d}

We can readily generalize our results to two-dimensional spacetime.
We work in the $\mathcal N=2$ superspace, parameterized by a set of holomorphic and anti-holomorphic coordinates:
\begin{equation}
    \left(z, \theta, \tilde{\theta}\right), \qquad \left( \bar{z}, \bar{\theta}, \bar{\tilde{\theta}} \right).
    \label{2d_coord_def}
\end{equation}
The two-dimensional superconformal group is a product of two one-dimensional superconformal groups for the left- and right-moving modes.
In particular, the $\mathcal N=2$ superconformal symmetry is realized by the $su(1,1|1) \oplus su(1,1|1)$ superalgebra.
As in one dimension, here we can use the superconformal symmetry to make the correlators depend only on bosonic coordinates.

The superalgebra has two commuting Casimir operators which are complex conjugates of each other.
We can write them in terms of bosonic cross-ratios as differential operators: 

\begin{equation}
    \mathcal C=\chi^2\left( 1-\chi \right)\partial_\chi^2+\chi\left( 1-\chi \right)\partial_\chi, \qquad \bar{\mathcal C}=\bar{\chi}^2\left( 1-\bar{\chi} \right)\partial_{\bar{\chi}}^2+\bar{\chi}\left( 1-\bar{\chi} \right)\partial_{\bar{\chi}},
    \label{C_2d}
\end{equation}
where $\chi, \bar{\chi}$ are holomorphic and anti-holomorphic cross-ratios:
\begin{equation}
   \chi\equiv \frac{z_{12}z_{34}}{z_{14}z_{32}}=\frac{\left \langle 12 \right \rangle \left \langle 34 \right \rangle }{\left \langle 14 \right \rangle \left \langle 32 \right \rangle }, \qquad \bar{\chi}\equiv \frac{\bar{z}_{12} \bar{z}_{34}}{\bar{z}_{14}\bar{z}_{32}}=\frac{\left \langle \bar{1}\bar{2} \right \rangle \left \langle \bar{3}\bar{4} \right \rangle }{\left \langle \bar{1}\bar{4} \right \rangle \left \langle \bar{3}\bar{2} \right \rangle }.
    \label{chi_c_def}
\end{equation}
Angle brackets $\left \langle ij \right \rangle $, $\left \langle \bar{i}\bar{j} \right \rangle$  denote the supersymmetric invariants, completely analogous to the ones we have seen in one dimension:
\begin{equation}
    \langle 12 \rangle = z_1-z_2-2\tilde{\theta}_1 \theta_2 - \theta_1 \tilde{\theta}_1 -\theta_2 \tilde{\theta}_2, \qquad \langle \bar{1}\bar{2} \rangle = \bar{z}_1 - \bar{z}_2-2\bar{\tilde{\theta}}_1\bar{\theta}_2-\bar{\theta}_1\bar{\tilde{\theta}}_1-\bar{\theta}_2 \bar{\tilde{\theta}}_2.
    \label{<>_2d_def}
\end{equation}
Knowing the eigenfunctions of the one-dimensional Casimir (\ref{C_phi}), we can easily guess the eigenfunctions and eigenvalues in two dimensions:
\begin{equation}
   \mathcal C \left(\varphi_h\left( \chi \right) \varphi_{\tilde{h}}\left( \bar{\chi} \right)\right)=h^2 \varphi_h\left( \chi \right) \varphi_{\tilde{h}}\left( \bar{\chi} \right), \qquad \bar{\mathcal C} \left(\varphi_h\left( \chi \right) \varphi_{\tilde{h}}\left( \bar{\chi} \right)\right)=\tilde{h}^2 \varphi_h\left( \chi \right) \varphi_{\tilde{h}}\left( \bar{\chi} \right).
   \label{C_2d_eigen}
\end{equation}
The $\varphi_h$ eigenfunction was defined in (\ref{phi_def}).
In what follows, we find a more convenient basis of the Casimir eigenfunctions using the shadow formalism.

On physical states, the Casimirs should be Hermitean conjugates, which gives us a condition:
\begin{equation}
   \overline{\left( h^2 \right)}=\tilde{h}^2 \qquad \Rightarrow \qquad \bar{h}=\tilde{h} \qquad \text{ or } \qquad \bar{h}=-\tilde{h}.
   \label{C-C_Herm}
\end{equation}
Another restriction comes from the fact that the spin of a bosonic physical state has to be real and in particular integer:
\begin{equation}
   l = h-\tilde{h} \in \mathbb Z,
   \label{J_def_integer}
\end{equation}
which implies that either spin is zero and both the dimensions $h=\tilde{h}$ are purely real, or the dimensions have the following form:
\begin{equation}
   h=\frac{l}{2}+is, \qquad \tilde{h} = -\frac{l}{2}+is, \qquad s \in \mathbb R.
   \label{h_h_form}
\end{equation}

To make the discussion more concrete, let's consider the $\mathcal N=2$ SYK model in two dimensions with complex scalar superfield and random superpotential.
Our goal is to find the conformal four-point function of the model:
\begin{equation}
   \mathcal W\left( \chi,\bar{\chi} \right)=\frac{\left \langle \tilde{\Phi}\left( 1,\bar{1} \right)\Phi \left( 2,\bar{2} \right)\tilde{\Phi} \left( 3,\bar{3} \right)\Phi\left( 4,\bar{4} \right) \right \rangle }{\left \langle \tilde{\Phi}\left( 1,\bar{1} \right)\Phi\left( 2,\bar{2} \right) \right \rangle \left \langle \tilde{\Phi}\left( 3,\bar{3} \right)\Phi\left( 4,\bar{4} \right) \right \rangle }.
   \label{4pt_2d}
\end{equation}
Here $\Phi$, $\tilde{\Phi}$ are chiral superfields with zero spin.
In a two-dimensional spacetime, a fermionic field  has scaling dimension $\frac12$, so a $q$-fermion interaction is generally irrelevant.
To make a $q$-particle interaction marginal, we consider scalar fields which have zero scaling dimension in the UV.
The chiral superfields are annihilated by superderivatives,
\begin{equation}
   {D}\tilde{\Phi}=\bar{D}\tilde{\Phi}=0,
   \label{D_Phi}
\end{equation}
defined as:
\begin{equation}
   D=\frac{\partial}{\partial \theta}+\tilde{\theta}\frac{\partial}{\partial z}, \qquad \bar{D}=\frac{\partial}{\partial \bar{\theta}}+\bar{\tilde{\theta}}\frac{\partial}{\partial z}.
   \label{D_2d_def}
\end{equation}

The Lagrangian of the model consists of a kinetic $D$-term and a superpotential $F$-term with random coupling:

\begin{equation}
   \mathcal L=\int  d^2\theta d^2\tilde{\theta} \Phi \tilde{\Phi}+ i\int  d^2\theta  C_{i_1 i_2 \dots i_{\hat{q}}}\Phi_{i_1} \dots \Phi_{i_{\hat{q}}}+ i\int  d^2\tilde{\theta}  \bar{C}_{i_1 i_2 \dots i_{\hat{q}}}\tilde{\Phi}_{i_1} \dots \tilde{\Phi}_{i_{\hat{q}}}, \qquad d^2\theta\equiv d\theta d\bar{\theta}.
    \label{L_2d}
\end{equation}
Here $\hat{q}$ can be any integer, and $C$ is a Gaussian coupling:
\begin{equation}
   \left \langle  C_{i_1 \dots i_{\hat{q}}}\bar{C}_{i_1 \dots i_{\hat{q}}}\right \rangle = \left( \hat{q}-1 \right)!\frac{J}{N^{\hat{q}-1}}.
   \label{C_2d_def}
\end{equation}
We assume that the $F$-term is not renormalized, perturbatively or non-perturbatively \cite{Vafa:1988uu}.
As an $\mathcal N=2$ superconformal theory with a holomorphic superpotential, we expect this model to flow to a conformal fixed point in the infrared.
The $D$-term gets renormalized and becomes irrelevant, so the infrared behavior of the model is determined exclusively by the superpotential.

Next we follow the same steps as for the one-dimensional model, finding first the two-point function, then the basis of the four-point functions in the shadow representation and finally eigenvalues of the kernel.

\subsection{Two-point function in two dimensions}

First we look for the chiral--anti-chiral two-point function:
\begin{equation}
   \mathcal G\left( 1|2 \right)\equiv\langle \tilde{\Phi}\left( 1, \bar{1}\right)\Phi\left( 2,\bar{2} \right)\rangle.
   \label{G_2d_def}
\end{equation}
The Lagrangian (\ref{L_2d}) implies the supersymmetric Schwinger--Dyson equation:
\begin{equation}
   D_1 \bar{D}_1 \mathcal G\left( 1|3 \right)+J\int d^2z_2 d^2\theta_2 \mathcal G\left(1|2  \right)\mathcal G^{\hat{q}-1}\left( 3|2 \right)=\left( \tilde{\theta}_1-\tilde{\theta}_3 \right)\left( \bar{\tilde{\theta}}_1-\bar{\tilde{\theta}}_3 \right)\delta\left( \left \langle 13 \right \rangle  \right)\delta\left( \left \langle \bar{1}\bar{3} \right \rangle  \right).
    \label{SD_2d}
\end{equation}

The $D^2\mathcal G$ term in the Schwinger--Dyson equation (\ref{SD_2d}) comes from differentiating the $D$-term.
In the usual non-supersymmetric SYK model, the conformal limit is identified with the large coupling limit, so in the conformal point we can neglect such a term.
When considering corrections to the conformal limit however, we have to restore it, and it gives a correction to the two-point function of order $\left( \beta J \right)^{-1}$.

In our case, the infrared behavior of the model should be completely determined by the superpotential, therefore the $D$-term should not affect the Schwinger--Dyson equation.
Hence we expect the integral equation (\ref{SD_2d}) to be true without the first term in the exact conformal limit. 

It is easy to see that the Schwinger--Dyson equation without the first term is satisfied by a conformal propagator of the form:
\begin{equation}
    \mathcal G\left( 1|2 \right)=\frac{b}{\langle 12 \rangle^{\Delta} \langle \bar{1}\bar{2} \rangle^\Delta}.
    \label{G_form_2d}
\end{equation}
Here $\Delta$ is the scaling dimension of the superconformal primary $\Phi$.
Dimensional considerations allow us to fix it:
\begin{equation}
    \hat{q}\Delta=1.
    \label{Delta_2d}
\end{equation}
The integral in (\ref{SD_2d}) can be taken in the momentum space.
We use the ansatz (\ref{G_form_2d}), integrate over the odd variables, and doing the Fourier transformation of the propagators with the help of an integral:
\begin{equation}
   \int \frac{d^2z}{|z|^{2\Delta}}e^{ip\cdot z}=|p|^{2\Delta-2}\cdot \frac{\pi}{2^{2\Delta-2}}\frac{\Gamma\left( 1-\Delta \right)}{\Gamma\left( \Delta \right)}.
    \label{G_fourier}
\end{equation}
Then the ansatz for the propagator works if we fix the $b$ constant to:
\begin{equation}
    b^{\hat{q}}J=\frac{1}{4\pi^2}.
    \label{b_2d}
\end{equation}

\subsection{Eigenfunctions of the Casimir operators}

Next we proceed to find the basis for the four-point function.
Just as in one-dimension, the eigenfunctions of the kernel can be found in the shadow representation.
These eigenfunctions are labeled by the eigenvalues of the Casimirs $\left( h,\tilde{h} \right)$.
We formally add an interaction term for fictitious superoperators $\mathcal V_{h,\tilde{h}}$:
\begin{equation}
   \varepsilon \int d^2z_0  d^2\theta_0  d^2\tilde{\theta}_0 \mathcal V_{h,\tilde{h}}\left( 0,\bar{0} \right)\mathcal V'_{-h,-\tilde{h}}\left( 0,\bar{0} \right).
   \label{shadow_term_2d}
\end{equation}
Note that here we integrate over the full superspace, i.e. this is a $D$-term.
The Casimir eigenfunction is given by an integral:

\begin{equation}
   \mathcal F_{h,\tilde{h}}\sim \int d^2z_0d^2\theta_0 d^2 \tilde{\theta}_0 \frac{\langle \tilde{\Phi}\left( 1,\bar{1} \right) \Phi\left( 2,\bar{2} \right) \mathcal V_{h,\tilde{h}}\left( 0,\bar{0} \right) \rangle  \langle \tilde{\Phi}\left( 3,\bar{3} \right) \Phi\left( 4 ,\bar{4}\right) \mathcal V'_{-h,-\tilde{h}}\left( 0,\bar{0} \right) \rangle}{\mathcal G(1|2) \mathcal G(3|4) }.
    \label{F_shadow_2d}
\end{equation}

The interaction term (\ref{shadow_term_2d}) makes it clear that eigenfunctions should remain invariant if we reverse the signs of both holomorphic and anti-holomorphic dimensions $\left( h,\tilde{h} \right)\leftrightarrow \left( -h,-\tilde{h} \right)$:
\begin{equation}
   \mathcal F_{h,\tilde{h}}=\mathcal F_{-h,-\tilde{h}}.
   \label{F_inv}
\end{equation}

Unlike in one dimension, here we can fix the three-point function uniquely, as a product of a holomorphic and an anti-holomorphic parts:
\begin{equation}
   \langle \tilde{\Phi}\left( 1,\bar{1} \right) \Phi\left( 2,\bar{2} \right) \mathcal V_{h,\tilde{h}}\left( 0,\bar{0} \right)\rangle = \frac{1}{\langle 12 \rangle^{\Delta-h}\langle 02 \rangle^h\langle 10 \rangle^h }\frac{1}{\langle \bar{1}\bar{2} \rangle^{\Delta-\tilde{h}}\langle \bar{0}\bar{2} \rangle^{\tilde{h}}\langle \bar{1}\bar{0} \rangle^{\tilde{h}} }.
    \label{3pt_holo}
\end{equation}
These three-point functions diagonalize both Casimirs $\mathcal C, \bar{\mathcal C}$, with eigenvalues $h^2, \tilde{h}^2$ correspondingly.
Dividing by propagators and integrating over the odd coordinates, we find the conformal block for the four-point function in the integral form, similar to (\ref{F_y}):

\begin{equation}
   \Xi_{h,\tilde{h}}=\left( -1 \right)^{h+\tilde{h}}\int dyd\bar{y} \frac{h \chi^h (1-y)^h}{y^h (\chi-y)^h}\left( \frac 1{y} + \frac{1}{\chi-y} -\frac{1}{1-y}\right) \frac{\tilde{h} \bar{\chi}^{\tilde{h}} (1-\bar{y})^{\tilde{h}}}{\bar{y}^{\tilde{h}} (\bar{\chi}-\bar{y})^{\tilde{h}}}\left( \frac 1{\bar{y}} + \frac{1}{\bar{\chi}-\bar{y}} -\frac{1}{1-\bar{y}}\right).
    \label{Xi_2d_int}
\end{equation}
Here we have added a $\left( -1 \right)^{h+\tilde{h}}$ factor to make our later expressions somewhat simpler. 
Just as in one dimension, here we see that the $\mathcal N=2$ four-point function does not depend on odd variables, unlike the $\mathcal N=1$ four-point function discussed in \cite{Murugan:2017eto}.

The integral (\ref{Xi_2d_int}) is tricky, but luckily we can use the results of \cite{Murugan:2017eto} for a two-dimensional bosonic SYK model.
The eigenbasis of the non-supersymmetric conformal Casimirs consists of the  $\Psi_{h,\tilde{h}}$ functions, where:

\begin{equation}
   \Psi_{h,\tilde{h}}\left( \chi,\bar{\chi} \right) \equiv \int dy d\bar{y} \frac{\chi^h (1-y)^{h-1}}{y^h \left( y-\chi \right)^h}\frac{\bar{\chi}^{\tilde{h}}\left( 1-y \right)^{\tilde{h}-1}}{\bar{y}^{\tilde{h}}\left(\bar{y}- \bar{\chi} \right)^{\tilde{h}}}.
    \label{Psi_2d_int}
\end{equation}
Explicitly, $\Psi_{h,\tilde{h}}$ is a combination of the eigenfunctions $F_h\left( \chi \right)$ (\ref{Fh_def}) of the non-supersymmetric one-dimensional conformal Casimir:
\begin{equation}
    \Psi_{h,\tilde{h}}\left( \chi,\bar{\chi} \right)=\frac12 \frac{\sin \pi h}{\cos \pi \tilde{h}}\left( F_h\left( \chi \right)F_{\tilde{h}}\left( \bar{\chi} \right)-F_{1-h}\left( \chi \right)F_{1-\tilde{h}}\left( \bar{\chi} \right) \right).
    \label{Psi_hh_Fh}
\end{equation}
Comparing the integral (\ref{Xi_2d_int}) with the definition of $\Psi_{h,\tilde{h}}$ (\ref{Psi_2d_int}), we see that the $\mathcal N=2$ eigenfunction is a linear combination of $\mathcal N=0$ eigenfunctions:
\begin{equation}
   \Xi_{h,\tilde{h}}=h\tilde{h}\left( \Psi_{h+1, \tilde{h}+1}+\Psi_{h,\tilde{h}}+\Psi_{h+1,\tilde{h}}+\Psi_{h,\tilde{h}+1} \right).
    \label{Xi_Psi_2d}
\end{equation}
The $\Xi_{h,\tilde{h}}$ eigenfunction is also a linear combination of the Casimir eigenfunctions (\ref{C_2d_eigen}):
\begin{equation}
   \Xi_{h,\tilde{h}}\left( \chi,\bar{\chi} \right)=h\tilde{h} \frac12 \frac{\sin \pi h}{\cos \pi \tilde{h}}\left( \varphi_h(\chi) \varphi_{\tilde{h}}\left( \bar{\chi} \right)-\varphi_{-h}(\chi)\varphi_{-\tilde{h}}\left( \bar{\chi} \right) \right).
    \label{Xi_phi_h}
\end{equation}

The eigenvalues of the Casimirs are:
\begin{equation}
   \mathcal C \Xi_{h,\tilde{h}}=h^2 \Xi_{h,\tilde{h}}, \qquad \bar{\mathcal C}\Xi_{h,\tilde{h}}=\tilde{h}^2 \Xi_{h,\tilde{h}}.
   \label{C_h^2}
\end{equation}
From this, it is clear that the spectrum of the Casimirs is symmetric under sign reversal:
\begin{equation}
    \Xi_{-h,-\tilde{h}}=\Xi_{h,\tilde{h}}.
    \label{Xi_-h}
\end{equation}

\subsection{Two-dimensional kernel}
\label{sec:kernel_2d}
The next step is to diagonalize the two-dimensional SYK kernel.
The $\mathcal N=2$ kernel is given by the same diagram (\ref{fig:kernel}) as before, and it reads as follows:
\begin{equation}
    K\left( 1',2'|1,2 \right)=\left( \hat{q}-1 \right)b^{\hat{q}}J\frac{1}{|\langle 11'\rangle|^{2\Delta}|\langle 2'2\rangle|^{2\Delta}|\langle 12\rangle|^{2-4\Delta} } d^2\tilde{\theta}_1  d^2\theta_2  d^2z_1 d^2z_2.
    \label{K_2d_def}
\end{equation}
Note that here, as well as in the one-dimensional case, we integrate only over half of the odd variables.

The kernel acts on the three-point function (\ref{3pt_holo}).
To simplify the calculations, we can take the coordinate of the $\mathcal V_{h,\tilde{h}}$ field to infinity, so that the three-point function becomes:
\begin{equation}
    f\left( 1,2,\infty; \bar{1}, \bar{2}, \infty \right)=\frac{1}{\langle 12 \rangle^{\Delta-h} \langle \bar{1}\bar{2} \rangle^{\Delta-\tilde{h}}}.
    \label{f_2d_limit}
\end{equation}
We can also conveniently fix the coordinates of the $1$ and $2$ points to be:

\begin{equation}
    1\to\left( 0,\tilde{{\vartheta}}_1,\bar{\tilde{\vartheta}}_1  \right), \qquad 2 \to \left( 1, {\vartheta}_2, \bar{{\vartheta}}_2 \right),
    \label{2d_fix}
\end{equation}
(the rest of the odd coordinates being zero) so that the corresponding invariants simplify:
\begin{equation}
    \langle 11' \rangle=z_1-\theta_1 \tilde{\theta}_1, \qquad \langle 2'2 \rangle=1-z_2-\theta_2 \tilde{\theta}_2, \qquad \langle 2'1' \rangle = 1.
    \label{<>_2d_simple}
\end{equation}
Then the eigenvalue of the kernel is:
\begin{equation}
    k\left( h,\tilde{h} \right)=\int K\left(\left. 1',2' \right|1,2\right) f\left( 1,2,\infty; \bar{1}, \bar{2}, \infty \right)=\frac{1-\Delta}{4\pi^2\Delta}\int\frac{\langle 12 \rangle^h \langle \bar{1}\bar{2} \rangle^{\tilde{h}}}{|\langle 11'\rangle|^{2\Delta}|\langle 2'2\rangle|^{2\Delta}|\langle 12\rangle|^{2-2\Delta} } d^2\tilde{\theta}_1  d^2\theta_2  d^2z_1 d^2z_2.
    \label{k_2d_def}
\end{equation}
In the integral over the odd variables, a non-zero contribution comes from the term containing $\tilde{\theta}_1  \bar{\tilde{\theta}}_1 \theta_2\bar{\theta}_2$.
It comes from the expansion of $\langle 12 \rangle^{h+\Delta-1}$ and $\langle \bar{1} \bar{2} \rangle^{\tilde{h}+\Delta-1}$.
Then after the integration, the eigenvalue becomes:
\begin{equation}
    k\left( h,\tilde{h} \right)=-\frac{\left(1-\Delta \right)}{\pi^2 \Delta} \left( -1+h+\Delta \right)\left( -1+\tilde{h}+\Delta \right)\int d^2z_1d^2z_2 \frac{\left( z_1-z_2 \right)^h \left( \bar{z}_1-\bar{z}_2 \right)^{\tilde{h}}}{|z_1|^{2\Delta}|z_2-1|^{2\Delta}|z_1-z_2|^{4-2\Delta}}.
    \label{k_2d_int}
\end{equation}
This expression can be evaluated explicitly with the help of the KLT integral (the calculation is completely analogous to what we did in Appendix \ref{sec:app_kernel} for the one-dimensional case):
\begin{equation}
    \int d^2x x^a \bar{x}^{\tilde{a}}\left( 1-x \right)^b \left( 1-\bar{x} \right)^{\tilde{b}}=\frac{\pi}{-1-a-b}\frac{B\left( 1+\tilde{a},1+\tilde{b} \right)}{B\left( -a,-b \right)},
    \label{KLT_int}
\end{equation}
the final answer being:
\begin{equation}
    k\left( h,\tilde{h} \right)=\Delta \left( 1-\Delta\right)\frac{\Gamma^2\left( -\Delta \right)}{\Gamma^2\left( \Delta \right)}\frac{\Gamma\left( -h+\Delta \right)\Gamma\left( \tilde{h}+\Delta \right)}{\Gamma\left( 1-h-\Delta \right)\Gamma\left( 1+\tilde{h}-\Delta \right)}.
    \label{k_2d_ans}
\end{equation}
This is the same as $k^{BB}$ in the $\mathcal N=1$ case \cite{Murugan:2017eto}, up to a sign:
\begin{equation}
    k\left( h,\tilde{h} \right)=-k^{BB}\left( h,\tilde{h} \right).
    \label{k_kBB}
\end{equation}

This eigenvalue has to be symmetric under $h \leftrightarrow \tilde{h}$, and it is if we restrict to physical states with either both dimensions real, or dimensions of the form (\ref{h_h_form}).
Also, for physical states the eigenvalue of the kernel is real.
So the condition on the operator spectrum $k\left( h,\tilde{h} \right)=1$ is a single real condition, therefore it is satisfied by a finite number of states for each spin.

As a check to our formula, we notice that there is a solution for $(h,\tilde{h})=(1,0)$, which corresponds to the $\mathcal N=2$ multiplet of the holomorphic superconformal current:
\begin{equation}
    \mathcal J=R+\theta S + \tilde{{\theta}}\tilde{S}+\theta \tilde{\theta}T,
    \label{J_def}
\end{equation}
which contains $R$-charge, supercurrent and stress tensor.
But unlike in one dimension, here the mode corresponding to the supercurrent is not in the Hilbert space (because neither of the conditions (\ref{C-C_Herm}) holds for the supercurrent), so it does not give rise to a divergence in the four-point function.

\subsection{Normalizable states and the full four-point function}
\label{sec:norm_2d}

As in the one-dimensional case, the next step towards finding the four-point function is to compute the norm of a state.
The inner product has to be invariant under the superconformal group, and the two-dimensional Casimir operators have to be Hermitean with respect to it.
Following the same logic as in Section \ref{sec:norm_1d}, we define the inner product as:
\begin{equation}
    \langle f\left( \chi,\bar{\chi} \right), g\left( \chi,\bar{\chi} \right) \rangle = \int \frac{d^2\chi}{|\chi|^2|1-\chi|^2}\bar{f}\left( \chi,\bar{\chi} \right)g\left( \chi,\bar{\chi} \right).
    \label{norm_2d}
\end{equation}
Unlike the one-dimensional inner product (\ref{fg_norm_red}), this one is real and the whole inner product is Hermitian.
Therefore we expect the eigenfunctions of the Casimir to form a usual Hilbert space, and be a complete set of functions (subject to a boundary condition analogous to (\ref{C_herm})). 

We expect the norm of an eigenfunction $\Xi_{h,\tilde{h}}$ to be proportional to $\delta$-function of a combination of $\left(h,\tilde{h}\right)$.
This singular contribution comes from the vicinity of zero.
Near $\chi \sim 0$, the eigenfunction behaves as a power of $\chi$:
\begin{equation}
    \Xi_{h,\tilde{h}}\left( \chi \right)\sim h\tilde{h} \frac{\sin \pi h}{2\cos \pi \tilde{h}}\left(B\left( h,h \right) B\left( \tilde{h}, \tilde{h} \right)\chi^h \bar{\chi}^{\tilde{h}}-B\left(-h,-h \right) B\left( -\tilde{h},- \tilde{h} \right)\chi^{-h}\bar{\chi}^{-\tilde{h}}\right), \qquad \chi \sim 0.
    \label{Xi_0}
\end{equation}
It is convenient to make a change of variables:
\begin{equation}
    \chi=e^{\rho+i\varphi}, \qquad \bar{\chi}=e^{\rho-i\varphi}.
    \label{chi_rho_phi}
\end{equation}
In these variables and near zero, the integration measure in (\ref{norm_2d}) becomes:
\begin{equation}
    \frac{d^2\chi}{|\chi|^2 |1-\chi|^2} \to d\rho d\varphi, \qquad \rho \to -\infty,
    \label{drho_dphi}
\end{equation}
and the eigenfunction is:
\begin{equation}
   \Xi_{h,\tilde{h}}\left( \chi \right)\sim h\tilde{h}\frac{\sin \pi h}{2\cos \pi \tilde{h}}\left(B\left( h,h \right) B\left( \tilde{h}, \tilde{h} \right)e^{\rho\left( h+\tilde{h} \right)+i\varphi \left( h-\tilde{h} \right)}-B\left(-h,-h \right) B\left( -\tilde{h},- \tilde{h} \right)e^{-\rho \left( h+\tilde{h} \right)-i\varphi\left( h-\tilde{h} \right)}\right).
    \label{Xi_0_rho}
\end{equation}
To make this function single-valued, we have to restrict the difference between eigenvalues to be integer:
\begin{equation}
    l\equiv h - \tilde{h}\in \mathbb Z.
    \label{l_def}
\end{equation}
This is natural since the operator $\mathcal V_{h,\tilde{h}}$ in the shadow representation has a bosonic lower component, and $l$ is its spin.
In particular, this means that we take the $\mathcal N=0$ eigenfunctions $\Psi_{h,\tilde{h}}$ which can be either even or odd under $\chi \to \frac{\chi}{\chi-1}$:
\begin{equation}
   \Psi_{h,\tilde{h}}\left( \frac{\chi}{\chi-1}, \frac{\bar{\chi}}{\bar{\chi}-1} \right)=\left( -1 \right)^{h-\tilde{h}}\Psi_{h,\tilde{h}}\left( \chi,\bar{\chi} \right).
    \label{Psi_chi_to}
\end{equation}
This is in contrast with the non-supersymmetric case, where $\chi \to \frac{\chi}{\chi-1}$ is a symmetry of the model and therefore the eigenfunction is even under this transformation. 
In our case, spin can be odd as well as even.
As in the one-dimensional case, the full $\mathcal N=2$ eigenfunction $\Xi_{h,\tilde{h}}$ is neither even nor odd under the $\chi \to \frac{\chi}{\chi-1}$ transformation, as is clear from (\ref{Xi_Psi_2d}).

We have seen in (\ref{C-C_Herm}) that the dimensions of the states in the Hilbert space have to either both be real,
\begin{equation}
   h=\tilde{h}\in \mathbb R,
   \label{h_real}
\end{equation}
or be of the form:
\begin{equation}
   h=\frac l2+is, \qquad \tilde{h}=-\frac l2+is, \qquad s \in \mathbb R.
   \label{h_form}
\end{equation}
In the former case, the eigenfunction (\ref{Xi_0_rho}) always diverges near zero, and the state is not normalizable.
In the latter, the product of two states is proportional to a delta function as desired.
If we further denote:
\begin{equation}
    \mathcal A\left( l,s \right)\equiv h\tilde{h} \frac{\sin \pi h}{2\cos \pi \tilde{h}}B\left( h,h \right) B\left( \tilde{h},\tilde{h} \right),
    \label{s_def}
\end{equation}
then the product of two states is:
\begin{multline}
   \left \langle \Xi_{s',l'}, \Xi_{s,l}\right\rangle \sim \int_0^{2\pi}d\varphi \int_{-\infty}^{0} d\rho \left( \mathcal A\left( l',-s' \right)e^{-i\rho s'-i\varphi l'}+\mathcal A\left( -l',s' \right)e^{i\rho s'+i\varphi l'} \right)\\
    \left( \mathcal A \left( l,s \right)e^{i\rho s+i\varphi l}+\mathcal A \left( -l,-s \right)e^{-i\rho s-i\varphi l} \right),
    \label{<>_2d_phi_rho}
\end{multline}
which gives after integration:
\begin{multline}
   \left \langle \Xi_{s',l'}, \Xi_{s,l}\right\rangle \sim 2\pi^2 \delta_{ll'} \delta\left( s-s' \right)\left(\mathcal A\left( l,-s \right)\mathcal A\left( l,s \right)+\mathcal A\left( -l,s \right)\mathcal A \left(-l,-s  \right)\right)\\
    +2\pi^2 \delta_{l,-l'} \delta\left( s+s' \right)\left(\mathcal A\left( l,-s \right)\mathcal A\left( l,s \right)+\mathcal A\left(-l,s \right)\mathcal A \left(-l,-s  \right)\right).
    \label{<>_delta_delta}
\end{multline}
The second line in (\ref{<>_delta_delta}) reflects the symmetry of the model under $\left( h,\tilde{h} \right) \leftrightarrow \left( -h,-\tilde{h} \right)$.
Using once again the Beta function identity (\ref{B+B-}) and the fact that $\bar{h}=-\tilde{h}$, we finally arrive at:
\begin{equation}
   \left \langle \Xi_{s',l'}, \Xi_{s,l}\right\rangle =4\pi^4 \left( l^2+s^2 \right)\left( \delta_{ll'} \delta\left( s-s' \right)+\delta_{l,-l'} \delta\left( s+s' \right)\right).
    \label{<>_2d_final}
\end{equation}
The norm is real and positive-definite for real $s$ and integer $l$, as expected of a norm in a Hilbert space.

This inner product gives rise to a completeness relation:
\begin{equation}
   \sum_{l=-\infty}^{\infty} \int_0^\infty \frac{ds}{2\pi} \frac{1}{2\pi^3\left( l^2+s^2 \right)}\overline{\Xi}_{h,\tilde{h}}\left( \chi,\bar{\chi} \right)\Xi_{h,\tilde{h}}\left( \chi',\bar{\chi}' \right)=|\chi|^2 |1-\chi|^2 \delta^2\left( \chi-\chi' \right).
    \label{complete_2d}
\end{equation}
There is a double pole in this expression, since the norm of a state with $l=s=0$ vanishes. 
We avoid this pole by infinitesimally deforming the integration contour to avoid the origin, as in fig. \ref{fig:contour_2}.

\subsection{Four-point function in two dimensions}
\label{sec:4pt_2d}
As the $\Xi_{h,\tilde{h}}$ eigenfunctions form a basis, we can find the full four-point function as an expansion:
\begin{equation}
    \mathcal F = \frac{1}{1-K}\mathcal F_0=\sum_{h,\tilde{h}} \frac{1}{1-k\left( h,\tilde{h} \right)}\frac{\langle \Xi_{h,\tilde{h}}, \mathcal F_0\rangle}{\langle \Xi_{h,\tilde{h}}, \Xi_{h,\tilde{h}}\rangle}\Xi_{h,\tilde{h}}.
    \label{F_2d_series}
\end{equation}
Here $\mathcal F_0$ is the zero-rung four-point function:
\begin{equation}
    \mathcal F_0=\chi^\Delta \bar{\chi}^\Delta.
    \label{F0_2d}
\end{equation}
To make use of the expansion (\ref{F_2d_series}), we have to find the inner product between a Casimir eigenfunction and the zero-rung four-point function $\langle \Xi_{h,\tilde{h}}, \mathcal F_0 \rangle$.
We can simplify the integral by acting on the eigenfunction with the Casimirs:
\begin{equation}
    \langle \mathcal C \bar{\mathcal C}\Xi_{h,\tilde{h}}, |\chi|^{2\Delta}\rangle= \left( h \tilde{h} \right)^2\langle \Xi_{h,\tilde{h}}, |\chi|^{2\Delta}\rangle= \langle \Xi_{h,\tilde{h}},\mathcal C \bar{\mathcal C} |\chi|^{2\Delta}\rangle=\Delta^4 \int d^2\chi \Xi_{h,\tilde{h}}\left( \chi,\bar{\chi} \right)|\chi|^{2\Delta-2}.
    \label{2d_use_herm}
\end{equation}
This expression looks similar to the $\mathcal N=0$ inner product:
\begin{equation}
    \left( f,g \right)\equiv \int \frac{d^2\chi}{|\chi|^4}\bar{f}g.
    \label{2d_n0_norm_def}
\end{equation}
Since the eigenfunction $\Xi_{h,\tilde{h}}$ is a linear combination of the $\mathcal N=0$ eigenfunctions $\Psi_{h,\tilde{h}}$ (\ref{Xi_Psi_2d}), we can express the $\mathcal N=2$ inner product via the non-supersymmetric one:
\begin{equation}
   \langle \Xi_{h,\tilde{h}}, |\chi|^{2\Delta}\rangle=\frac{\Delta^4}{h\tilde{h}}\left( \left( \Psi_{h,\tilde{h}},|\chi|^{2\Delta+2} \right)+\left( \Psi_{h+1,\tilde{h}+1},|\chi|^{2\Delta+2} \right) +\left( \Psi_{h+1,\tilde{h}},|\chi|^{2\Delta+2} \right) +\left( \Psi_{h,\tilde{h}+1},|\chi|^{2\Delta+2} \right)\right).
    \label{<>_n2_<>_n0}
\end{equation}
Now we can apply the results of \cite{Murugan:2017eto} about the $\mathcal N=0$ inner product:
\begin{equation}
    \left( \Psi_{h,\tilde{h}},|\chi|^{2\Delta} \right)=\frac{\pi^2 \Delta}{\left( 2-\Delta \right)\left( 1-\Delta \right)^2}k_{\mathcal N=0}\left( h,\tilde{h} \right)=\frac{\pi^2 \Delta}{1-\Delta}\frac{k\left( h,\tilde{h} \right)}{\left( -1+h+\Delta \right)\left( -1+\tilde{h}+\Delta \right)},
    \label{<>_n0_1}
\end{equation}
where $k\left( h,\tilde{h} \right)$ is the eigenvalue of the $\mathcal N=2$ kernel (\ref{k_2d_ans}). Explicitly, it is:
\begin{equation}
    \left( \Psi_{h,\tilde{h}},|\chi|^{2\Delta} \right)=-\pi^2 \frac{\Gamma^2\left( 1-\Delta \right)}{\Gamma^2\left( \Delta \right)}\frac{\Gamma\left( -h+\Delta \right)\Gamma\left( \tilde{h}+\Delta-1 \right)}{\Gamma\left( 2-h-\Delta \right)\Gamma\left( \tilde{h}-\Delta+1\right)}.
    \label{n0_expl}
\end{equation}
Plugging this in the sum (\ref{<>_n2_<>_n0}), we finally get:
\begin{equation}
   \langle \Xi_{h,\tilde{h}}, \mathcal F_0 \rangle = \frac{4\pi^2 \Delta}{1-\Delta}k\left( h,\tilde{h} \right).
    \label{<Xi_F0>_ans}
\end{equation}
As in all versions of the SYK model we've been discussing so far, the inner product with the zero-rung four-point function is proportional to the eigenvalue of the kernel.

%

Using this answer in (\ref{F_2d_series}), together with the norm of an eigenfunction (\ref{<>_2d_final}), we write the full four-point function as follows:
\begin{equation}
   \mathcal F\left( \chi,\bar{\chi} \right)=-\frac{2}{\pi} \frac{\Delta}{1-\Delta}\sum_{l \in \mathbb Z}\int_{0}^\infty \frac{ds}{2\pi} \frac{1}{l^2+s^2}\frac{k\left( h,\tilde{h} \right)}{1-k\left( h,\tilde{h} \right)}\Xi_{h,\tilde{h}}\left( \chi,\bar{\chi} \right).
    \label{F_2d_ans}
\end{equation}
The symmetry of the eigenfunctions under $\left( h,\tilde{h} \right) \leftrightarrow \left( -h,-\tilde{h} \right)$ allows us to put it in the form:
\begin{equation}
   \mathcal F\left( \chi,\bar{\chi} \right)=\frac{1}{4\pi} \frac{\Delta}{1-\Delta}\sum_{l \in \mathbb Z}\int_{-\infty}^\infty \frac{ds}{2\pi} \frac{k\left( h,\tilde{h} \right)}{1-k\left( h,\tilde{h} \right)}\frac{\sin \pi h}{\cos \pi \tilde{h}}\varphi_h\left( \chi \right)\varphi_{\tilde{h}}\left( \bar{\chi} \right).
    \label{F_2d_ans+}
\end{equation}

From this, we can find the central charge of the model.
On general grounds, the central charge of an $\mathcal N=2$ two-dimensional CFT of $N$ superfields and with a superpotential of degree $q$ is \cite{Vafa:1988uu}:
\begin{equation}
   c=\sum_{i=1}^N 6\left( \frac12-\frac{1}{q} \right)=3N\left( 1-2\Delta \right).
   \label{c_N=2}
\end{equation}
Now let us confirm this central charge from the four-point function (\ref{F_2d_ans+}).
As was found in \cite{Murugan:2017eto}, the stress tensor contributes to the $\chi^2$ term of the four-point function, so this term depends on the central charge:
\begin{equation}
    \mathcal F = \dots + \frac{N \Delta^2}{2c}\chi^2+O\left( \chi^2 \right).
    \label{F_T}
\end{equation}
The stress tensor lives in the supercurrent multiplet, which is a $(1,0)$ primary. 
At $\left( h,\tilde{h} \right)=\left( 1,0 \right)$, or equivalently at $\left( l,s \right)=\left( 1, i \right)$ the integrand in (\ref{F_2d_ans}) has a pole.
Taking $h=1+\tilde{h}=1+\varepsilon$ and expanding everything in $\varepsilon$, we find:
\begin{eqnarray}
    \varphi_{\varepsilon}\left( \bar{\chi} \right)&=&  \frac{2}{\varepsilon}+O(\varepsilon),\\
    \varphi_{1+\varepsilon}\left( \chi \right)&=&  \chi+\frac{\chi^2}{3}+O(\varepsilon),\\
    k\left( 1+ \varepsilon,\varepsilon\right)&= &1+\frac{1-2\Delta}{\Delta\left( 1-\Delta \right)}\varepsilon+O\left( \varepsilon^2 \right).
    \label{phi_eps_exp}
\end{eqnarray}
(The expressions for $\varphi_h$ can be derived e.g. from (\ref{F_series}).)
Bringing everything together, we find the central charge:
\begin{equation}
   c=3N\left( 1-2\Delta \right).
   \label{c_ans}
\end{equation}

This is exactly twice the central charge of the $\mathcal N=1$ model found in \cite{Murugan:2017eto}:
\begin{equation}
   c_{\mathcal N=2}=2c_{\mathcal N=1}.
   \label{c2_c1}
\end{equation}

\subsection{Retarded kernel in two dimensions}
\label{sec:chaos_2d}
We can now generalize the analysis of Section \ref{sec:retarded} to the two-dimensional system, to find the chaos exponent and identify the modes contributing to it.
To do that, we construct the kernel out of retarded and left-right propagators (see fig. \ref{fig:retarded}).
We proceed in the same fashion as before, doing an analytical continuation and putting one rail of the ladder diagram at $\tau_l=it$ and the other at $\tau_r=it+\pi $. 
We also transform the coordinates from $\left( z, \theta, \tilde{\theta} \right)$ to the periodic $\left( w,\vartheta, \tilde{\vartheta}\right)$, where:
\begin{equation}
   w=x+i\tau=x-t, \qquad \bar{w}=x-i\tau=x+t.
   \label{w_per_def}
\end{equation}
The coordinate transformation differs for the left and the right rails:
\begin{equation}
    \begin{array}{llcrr}
       z_1=e^{w_1}&&\qquad& z_2=e^{w_2+i\pi}=-e^{w_2},&\\
        \theta_1=e^{\frac{w_1}{2}}\vartheta_1,&\text{(left rail)}&\qquad&\theta_2=e^{\frac{w_2+i\pi}{2}}=ie^{\frac{w_2}{2}}\vartheta_2,&\text{(right rail)}\\
        \tilde{\theta}_1=e^{\frac{w_1}{2}}\tilde{\vartheta}_1,&&\qquad&\tilde{\theta}_2=e^{\frac{w_2-i\pi}{2}}=-ie^{\frac{w_2}{2}}\tilde{\vartheta}_2.&
    \end{array}
    \label{coord_2d}
\end{equation}
To make the expressions more symmetrical, we take a different transformation for the anti-holomorphic coordinates:
\begin{equation}
    \begin{array}{llcrr}
       \bar{z}_1=e^{-\bar{w}_1}&&\qquad& z_2=e^{-\bar{w}_2+i\pi}=-e^{-\bar{w}_2},&\\
       \bar{\theta}_1=e^{-\frac{\bar{w}_1}{2}}\bar{\vartheta}_1,&\text{(left rail)}&\qquad&\bar{\theta}_2=e^{\frac{-\bar{w}_2+i\pi}{2}}=ie^{-\frac{\bar{w}_2}{2}}\bar{\vartheta}_2,&\text{(right rail)}\\
       \bar{\tilde{\theta}}_1=e^{-\frac{\bar{w}_1}{2}}\bar{\tilde{\vartheta}}_1,&&\qquad&\bar{\tilde{\theta}}_2=e^{-\frac{\bar{w}_2-i\pi}{2}}=-ie^{-\frac{\bar{w}_2}{2}}\bar{\tilde{\vartheta}}_2.&
    \end{array}
    \label{coord_2d_antiholo}
\end{equation}
Then the supersymmetry-invariant distance between two points belonging to the same rail is:
\begin{equation}
   \langle 11' \rangle_{ll}=e^{\frac{w_1+w_{1'}}2}\left( 2\sinh  \frac{w_1-w_{1'}}{2} -2\tilde{\vartheta}_1 \vartheta_{1'}-\vartheta_1 \tilde{\vartheta}_{1'}-\vartheta_{1'} \tilde{\vartheta}_{1'}\right),
   \label{11'_ret_holo}
\end{equation}
and the invariant distance between the rails is:
\begin{equation}
   \langle 12 \rangle_{lr}=e^{\frac{w_1+w_2}{2}}\left( 2 \cosh  \frac{w_1-w_2}{2} -2i\tilde{\vartheta}_1 \vartheta_2 - e^{\frac{w_1-w_2}2}\vartheta_1 \tilde{\vartheta}_1 -e^{\frac{w_2-w_1}{2}}\vartheta_2 \tilde{\vartheta}_2\right).
   \label{12'_ret_holo}
\end{equation}
For the anti-holomorphic invariants, the exponents in (\ref{11'_ret_holo}, \ref{12'_ret_holo}) are negative:
\begin{eqnarray}
   \label{11'_ret_antiholo}
   \langle \bar{1}\bar{1}' \rangle_{ll}&=&e^{-\frac{\bar{w}_1+\bar{w}_{1'}}2}\left( 2\sinh  \frac{\bar{w}_1-\bar{w}_{1'}}{2} -2\bar{\tilde{\vartheta}}_1 \bar{\vartheta}_{1'}-\bar{\vartheta}_1 \bar{\tilde{\vartheta}}_{1'}-\bar{\vartheta}_{1'} \bar{\tilde{\vartheta}}_{1'}\right),\\
   \langle \bar{1}\bar{2} \rangle_{lr}&=&e^{-\frac{\bar{w}_1+\bar{w}_2}{2}}\left( 2 \cosh  \frac{\bar{w}_1-\bar{w}_2}{2} -2i\bar{\tilde{\vartheta}}_1 \bar{\vartheta}_2 - e^{\frac{\bar{w}_1-\bar{w}_2}2}\bar{\vartheta}_1 \bar{\tilde{\vartheta}}_1 -e^{\frac{\bar{w}_2-\bar{w}_1}{2}}\bar{\vartheta}_2 \bar{\tilde{\vartheta}}_2\right).
   \label{12'_ret_antiholo}
\end{eqnarray}

Knowing these supersymmetric invariants, we can construct retarded propagators.
To do that, we once again add an infinitesimal imaginary part to $t$,
\begin{equation}
   t \to t \pm i\epsilon,
   \label{t_2d_eps}
\end{equation}
and compute the difference:
\begin{equation}
   \mathcal G_R\left( 1|1' \right)=\Theta\left( t_1-t_{1'} \right)\left( \mathcal G\left(w_1+i\epsilon,\bar{w}_1-i\epsilon|w_{1'}, \bar{w}_{1'} \right)-\mathcal G\left(w_1-i\epsilon,\bar{w}_1+i\epsilon|w_{1'}, \bar{w}_{1'} \right)  \right),
   \label{G_R_2d_def}
\end{equation}
where we have omitted the Grassmann coordinates for brevity.
Taking into account the Jacobian of the transformation, we find:
\begin{equation}
   \mathcal G_R\left( 1|1' \right)=\Theta\left( t_{11'}-|x_{11'}| \right)\frac{-2ib \sin \pi \Delta}{ \langle 11^{'} \rangle_{ll}^\Delta \langle \bar{1}\bar{1}'\rangle_{ll}^\Delta}e^{\frac{\Delta}{2}\left( w_1-\bar{w}_1 \right)}e^{\frac{\Delta}{2}\left( w_{1'}-\bar{w}_{1'} \right)}.
   \label{GR_2d_expl}
\end{equation}
The left-right propagator is simply:
\begin{equation}
   \mathcal G_{lr}\left( 1|2 \right)=\frac{b}{\langle 12 \rangle_{lr}^\Delta \langle \bar{1}\bar{2}\rangle_{lr}^{\Delta}}e^{\frac{\Delta}{2}\left( w_1-\bar{w}_1 \right)}e^{\frac{\Delta}{2}\left( w_2-\bar{w}_2 \right)}.
   \label{G_lr}
\end{equation}
From these propagators, we can build the two-dimensional retarded kernel:
\begin{equation}
K_R\left( 1,2|1',2' \right)=\frac14 J \left( \hat{q}-1 \right)\mathcal G_R\left( 1'|1 \right)\mathcal G_R\left( 2|2' \right) \mathcal G_{lr}^{\hat{q}-2}\left( 1'|2' \right)e^{\frac12\left( w_{1'}-\bar{w}_{1'} \right)}e^{\frac12\left( w_{2'}-\bar{w}_{2'} \right)} d^2w_{1'}d^2w_{2'}d\tilde{\vartheta}_{1'}d\bar{\tilde{\vartheta}}_{1'}d\vartheta_{2'}d\bar{\vartheta}_{2'}.
   \label{KR_2d_def}
\end{equation}
Using the explicit form of the propagators, we find for the kernel:
\begin{multline}
   K_R\left( 1,2|1',2' \right)=\sin^2 \pi \Delta \left( Jb^{\hat{q}} \right) \left( \hat{q}-1 \right)e^{-\Delta\left( t_1+t_2 \right)}e^{\Delta\left( t_{1'}+t_{2'} \right)}\cdot \\
   \frac{\Theta\left( t_{11'}-|x_{11'}| \right)\Theta\left( t_{22'}-|x_{22'}| \right)}{\langle 1'1 \rangle_{ll}^\Delta \langle \bar{1}'\bar{1}\rangle_{ll}^\Delta\langle 22' \rangle_{ll}^\Delta \langle \bar{2}\bar{2}'\rangle_{ll}^\Delta\langle 1'2' \rangle_{lr}^{1-2\Delta} \langle \bar{1}'\bar{2}'\rangle_{lr}^{1-2\Delta}} d^2w_{1'}d^2w_{2'}d\tilde{\vartheta}_{1'}d\bar{\tilde{\vartheta}}_{1'}d\vartheta_{2'}d\bar{\vartheta}_{2'}.
   \label{KR_2d_int}
\end{multline}
This kernel is diagonalized by three-point functions. 
We take the coordinate of one of the operator insertions to infinity, and write the eigenfunction of (\ref{KR_2d_int}) as:
\begin{equation}
   f_R\left( 1,2,\infty \right)=e^{-\Delta\left( w_{1}+w_{2} \right)}e^{\Delta\left( \bar{w}_{1}+\bar{w}_{2} \right)}\frac{1}{\langle 12 \rangle_{lr}^{\Delta-h} \langle \bar{1}\bar{2}\rangle_{lr}^{\Delta-\tilde{h}}}.
   \label{fR_def_2d}
\end{equation}
To see if this three-point function grows with time, consider its bosonic part at $w_1=w_2$.
Then, using (\ref{12'_ret_holo}), we reduce the retarded three-point function to:
\begin{equation}
   f_R \sim e^{\left( h-\tilde{h} \right)x}e^{-\left( h+\tilde{h} \right)t}.
   \label{fR_growth}
\end{equation}
We want to find a mode which exhibits exponential growth in time, and no growth in space.
Therefore, we restrict:
\begin{equation}
   h-\tilde{h} \in i \mathbb R,
   \label{h-h_Im}
\end{equation}
and look for a mode with negative $h+\tilde{h}$ and the eigenvalue of the retarded kernel equal to one.

Fixing the variables in (\ref{fR_def_2d}) and (\ref{KR_2d_int}):
\begin{equation}
   1=\left( 0, \vartheta_1=\bar{\vartheta}_1=0 \right), \qquad 2=\left( 0, \tilde{\vartheta}_2=\bar{\tilde{\vartheta}}_2=0 \right),
   \label{2d_fix_ret}
\end{equation}
we write the eigenvalue of the kernel as an integral:
\begin{equation}
   k_R\left( h,\tilde{h} \right)=\int K\left( 1,2|1',2' \right)f_R\left(1',2',\infty  \right).
   \label{kR_2d_def}
\end{equation}
In this expression, the left- and right-moving modes are completely decoupled.
We can integrate over odd variables and then use the same integral as for the one-dimensional kernel (\ref{kr_int}) to find:
\begin{equation}
   k_R\left( h,\tilde{h} \right)=-\frac{\Gamma^2\left( 1-\Delta \right)}{\Gamma\left( \Delta+1 \right)\Gamma\left( \Delta-1 \right)}\frac{\Gamma\left( \Delta-h \right)\Gamma\left( \Delta-\tilde{h} \right)}{\Gamma\left( 1-\Delta-h \right)\Gamma\left( 1-\Delta-\tilde{h} \right)},
   \label{kR_2d_ans}
\end{equation}
which exactly coincides with the kernel of the $\mathcal N=1$ two-dimensional model.

If the difference $h-\tilde{h}$ is imaginary, this eigenvalue of the kernel is real. 
Therefore the $k_R=1$ condition has a continuous family of solutions for different $h,\tilde{h}$.
As has already been discussed in \cite{Murugan:2017eto}, the chaos exponents found in this model are below $\approx 0.6$, thus not saturating the maximal chaos bound.

%
\section{Conclusion}
In this paper, we present a technical computation of the four-point function of an SYK-inspired model with $\mathcal N=2$ symmetry.
We follow the outline of \cite{Maldacena:2016hyu}, finding first the eigenbasis of the superconformal Casimir and then the action of the SYK kernel on the eigenfunctions.
We find the two-particle Casimir of the $\mathcal N=2$ superconformal group as a differential operator (\ref{C_def}) and then compute its eigenfunctions, first directly solving the eigenvalue equation and then using the shadow representation.
Then we expand the four-point function of the $\mathcal N=2$ SYK model in this basis, with the result being (\ref{F_1d_full}).
We can also write the four-point function as a sum over the positive solutions to the $k(h)=1$ equation.

We find the $\mathcal N=2$ SYK model very similar to the non-supersymmetric model with complex fermions.
The eigenfunctions of the two-particle Casimir are linear combinations of the $\mathcal N=0$ eigenfunctions, and the supergroup can be used to make the four-point function depend only on the  bosonic coordinates.
The $\mathcal N=2$ eigenfunctions with the conformally invariant inner product do not form a Hilbert space, and the norm is positive semi-definite in that case.
Nevertheless, we can expand the zero-rung four-point function in the eigenfunctions of the Casimir and use this expansion to find the full four-point function.
This four-point function has a pole at $h=1$, which corresponds to the supercharge multiplet, containing the $R$-charge, the stress tensor and two supercharges. 
To resolve this pole, we would have to consider the model away from the conformal limit, which is beyond the scope of this paper.
A discussion of such a resolution can be found in \cite{Fu:2016vas}.
We also find that the $h=-1$ mode is maximally chaotic in the out-of-time order four-point function, just as in the non-supersymmetric case.

Since the two-dimensional $\mathcal N=2$ superalgebra is a direct sum of holomorphic and anti-holomorphic copies of one-dimensional $su(1,1|1)$ superalgebras, our results can be easily generalized to the two-dimensional space.
We consider a model containing chiral superfields with random holomorphic superpotential and find the expansion of the four-point function in terms of eigenfunctions of the two-dimensional Casimir (\ref{F_2d_ans}).
We also check that the equation $k(h)=1$ is satisfied for the supercurrent multiplet with $(h,\tilde{h})=(1,0)$.
The retarded kernel for this model exactly coincides with the one for the $\mathcal N=1$ two-dimensional SYK model, which has been found in \cite{Murugan:2017eto} to be non-maximally chaotic.
We also find the central charge of the $\mathcal N=2$ model to be twice that of an $\mathcal N=1$ model.

There are numerous broad questions one can ask about the $\mathcal N=2$ SYK model.
They include the existence of true RG fixed points outside the large $N$ limit; the realization of this model without random potential in spirit of \cite{Witten:2016iux}; a possible holographic dual or further extension to higher dimensions.
We hope to address some of these questions elsewhere.
\appendix
\section{\(\mathcal N=2\) Casimir}
\label{sec:app_Casimir}

The generators of the $SU(1,1|1)$ superconformal algebra can be presented in the differential form:

\begin{eqnarray}
    L_0&=& -\tau \partial_\tau -\frac12 \theta \partial_\theta -\frac12 \bar{\theta} \partial_{\bar{\theta}}-\Delta,\\
    L_1&=& - \partial_\tau,\\
    L_{-1}&=& -\tau^2 \partial_\tau -\tau \theta \partial_\theta -\tau \bar{\theta} \partial_{\bar{\theta}}-2\tau \Delta-\frac{Q}{2}\theta \bar{\theta},\\
    J_0&=& -\theta\partial_\theta+\bar{\theta}\partial_{\bar{\theta}}+Q,\\
    G_{+1/2}&=& \tau \partial_{\bar{\theta}}- \tau \theta \partial_\tau - \left( 2\Delta+ Q/2 \right)\theta- \theta \bar{\theta}\partial_{\bar{\theta}},\\
    G_{-1/2}&=&  \partial_{\bar{\theta}}- \theta \partial_\tau ,\\
    \bar{G}_{+1/2}&=&  \tau \partial_\theta- \tau\bar{\theta}\partial_\tau-\left( 2\Delta -Q/2 \right)\bar{\theta}+ \theta\bar{\theta}\partial_\theta ,\\
    \bar{G}_{-1/2}&=&  \partial_\theta-\bar{\theta}\partial_\tau .
    \label{N=2_gen}
\end{eqnarray}

A one-particle quadratic Casimir then is:

\begin{equation}
    C_2=L_0 L_0 - \frac14 J_0 J_0 -L_{1}L_{-1}+\frac12 G_{+1/2} \bar{G}_{-1/2}+\frac12 \bar{G}_{+1/2} G_{-1/2}. 
    \label{C_2_def}
\end{equation}

It commutes with all the other generators of the algebra.
It acts on bosonic functions as:

\begin{equation}
    C_2 f(\tau)=\left( \Delta^2 -\frac{Q^2}{4} \right)f(\tau),
    \label{C_tau}
\end{equation}
and on fermionic coordinates as:
\begin{eqnarray}
    C_2 \theta&=& \left( \Delta^2 -\frac{Q^2}{4}+\frac{Q}{4} \right)\theta,\\
    C_2 \bar{\theta}&=& \left( \Delta^2 -\frac{Q^2}{4}-\frac{Q}{4} \right)\bar{\theta}.
    \label{C_2_theta}
\end{eqnarray}

A two-particle operator is defined as a sum of one-particle operators:

\begin{equation}
    L^{2p}_0= L^{(1)}_0+L^{(2)}_{0},
    \label{2p_def}
\end{equation}
and so on.
The two-particle Casimir is the same expression (\ref{C_2_def}), written in terms of two-particle operators:

\begin{equation}
    C^{2p}=L^{2p}_0 L^{2p}_0 - \frac14 J^{2p}_0 J^{2p}_0 -L^{2p}_{1}L^{2p}_{-1}+\frac12 G^{2p}_{+1/2} \bar{G}^{2p}_{-1/2}+\frac12 \bar{G}^{2p}_{+1/2} G^{2p}_{-1/2}. 
    \label{C_2p_def}
\end{equation}

The Casimir acts on chiral-antichiral correlation functions, so we take the $R$-charge to be zero:
\begin{equation}
    Q=0.
    \label{Q=0}
\end{equation}

Then the eigenvalue of one-particle Casimir is $\Delta^2$.
The two-particle Casimir acts on the functions of the cross-ratio $\chi$, conjugated with a two-point function:

\begin{equation}
    C^{2p}\left( \frac{\sgn \tau_{12}}{|\langle 12 \rangle|^{2\Delta}}f\left( \chi \right) \right)=\frac{\sgn \tau_{12}}{|\langle 12 \rangle|^{2\Delta}}\mathcal C\left( \chi \right) f\left( \chi \right),
    \label{Cf}
\end{equation}
where $\mathcal C$ is a second-order differential operator:
\begin{equation}
    \mathcal C=\chi^2\left( 1-\chi \right)\partial_\chi^2+\chi\left( 1-\chi \right)\partial_\chi.
    \label{C_def_app}
\end{equation}

\section{\(\mathcal N=0\) SYK with complex fermions}
\label{sec:app_n0}
Here we list the eigenfunctions of the $\mathcal N=0$ Casimir.
In terms of the cross-ratio, the Casimir reads:

\begin{equation}
    \mathcal C^{\mathcal N=0}=\chi^2\left( 1-\chi \right)\partial^2_\chi-\chi^2\partial_\chi.
    \label{C_N=0_app}
\end{equation}
The eigenvalues of the Casimir are $h\left( h-1 \right)$ and the eigenfunctions $F_h, F_{1-h}$:

\begin{equation}
    F_h\left( \chi \right)\equiv \frac{\Gamma^2(h)}{\Gamma(2h)} \chi^h \,_2F_1\left( h,h;2h;\chi \right), \qquad \chi<1,
    \label{Fh_def}
\end{equation}

\begin{equation}
    \mathcal C^{\mathcal N=0}F_h=h\left( h-1 \right)F_h.
    \label{CFh}
\end{equation}

The eigenfunctions of the Casimir can be $\mathcal T$-even and $\mathcal T$-odd.
The $\mathcal T$-even eigenfunctions can be either anti-symmetric or symmetric under exchange of fermions.
Explicitly, they are:

\begin{equation}
    \Psi^{A}_h\left(\chi  \right)=\left\{
    \begin{gathered}
        \frac{2}{\cos \pi h}\left(  \cos^2 \frac{\pi h}{2}F_h(\chi) -  \sin^2 \frac{\pi h}{2} F_{1-h}(\chi)\right), \qquad \chi<1,\\
        \frac{2}{\sqrt{\pi}} \Gamma\left( \frac{h}{2} \right) \Gamma\left( \frac{1-h}{2} \right)\,_2F_1\left( \frac{h}{2},\frac{1-h}{2};\frac{1}{2}; \frac{\left( 2-\chi \right)^2}{\chi^2} \right), \qquad \chi>1.
    \end{gathered}
    \right.
    \label{IAA_2F1}
\end{equation}
and:

\begin{equation}
    \Psi^{S}_h\left(\chi  \right)=\left\{
        \begin{aligned}[r]
        \frac{2}{\cos \pi h}\left(   - \sin^2 \frac{\pi h}{2}F_h(\chi)+ \cos^2 \frac{\pi h}{2} F_{1-h}(\chi)\right), \qquad \chi<1,\\
        -\frac{4}{\sqrt{\pi}} \left( \frac{2-\chi}{\chi} \right)\Gamma\left( 1-\frac{h}{2} \right)\Gamma\left( \frac{1+h}{2} \right) \,_2F_1\left( 1-\frac{h}{2},\frac{1+h}{2};\frac{3}{2}; \frac{\left( 2-\chi \right)^2}{\chi^2} \right), \\
        \chi>1.
    \end{aligned}
    \right.
    \label{IBB_2F1}
\end{equation}

The $\mathcal T$-breaking eigenfunctions have mixed symmetry: they are odd under exchange of one pair of fermions and odd under exchange of the other. 
They can also be written in terms of $F_h$:

\begin{equation}
    \Psi^{AS}_h\left( \chi \right)=\left\{
        \begin{gathered}
            \frac{1}{\pi}\sin^2 \frac{\pi h}{2}\tan \pi h \left( F_h\left( \chi \right)-F_{1-h}\left( \chi \right) \right), \qquad \chi<1,\\
            0, \qquad \chi>1.
        \end{gathered}
        \right.
    \label{IAB_FG}
\end{equation}

\begin{equation}
    \Psi^{SA}_h\left( \chi \right)=\left\{
        \begin{gathered}
            \frac{1}{\pi}\cos^2 \frac{\pi h}{2}\tan \pi h \sgn\left( \chi \right)\left( F_h\left( \chi \right)-F_{1-h}\left( \chi \right) \right), \qquad \chi<1,\\
            0, \qquad \chi>1.
        \end{gathered}
        \right.
    \label{IBA_FG}
\end{equation}

The $\mathcal T$-even eigenfunctions have bound states.
The anti-symmetric eigenfunction is normalizable at even positive $h$, and the symmetric one is normalizable at odd positive $h$, with the spectrum of course being symmetric under $h \leftrightarrow 1-h$.

The eigenvalues of the kernel in non-supersymmetric model are also of two types:

\begin{equation}
    k^A_{\mathcal N=0}\left( h, \Delta\right)=\frac{1}{\pi} \frac{\Gamma\left( -2\Delta \right)}{\Gamma\left( 2\Delta -2\right)}\Gamma\left( 2\Delta-h \right)\Gamma\left( 2\Delta+h-1 \right)\left(\sin \pi h - \sin 2\pi \Delta \right).
    \label{kA_ans_n0}
\end{equation}

\begin{equation}
    k^S_{\mathcal N=0}\left( h,\Delta \right)=\frac{1}{\pi}\frac{\Gamma\left(1 -2\Delta \right)}{\Gamma\left( 2\Delta-1 \right)}\Gamma\left( 2\Delta-h \right)\Gamma\left( 2\Delta+h-1 \right)\left( \sin \pi h + \sin 2 \pi \Delta \right).
    \label{kB_ans_n0}
\end{equation}

\section{Eigenfunctions of the \(\mathcal N=0\) and \(\mathcal N=2\) superconformal Casimirs}
\label{sec:app_Fh_phi}
Here we show the relation between eigenfunctions:

\begin{equation}
    \varphi_h\left( \chi \right)=F_h\left( \chi \right)-F_{h+1}\left( \chi \right).
    \label{app_phi_Fh}
\end{equation}
Given the relation between the Casimir operators:
\begin{equation}
    \mathcal C_{\mathcal N=2}=\mathcal C_{\mathcal N=0}+\chi \partial_\chi,
    \label{app_C0_C2}
\end{equation}
we find that the $\mathcal N=2$ Casimir acts on the combination (\ref{app_phi_Fh}) as:
\begin{equation}
    \left( \mathcal C_{\mathcal N=0}+\chi \partial_\chi \right)\left( F_h-F_{h+1} \right)=h\left( h-1 \right) F_{h}-h\left( h+1 \right)F_{h+1}+\chi\partial_\chi\left( F_h-F_{h+1} \right)=h^2\left( F_h-F_{h+1} \right).
    \label{C_F-F}
\end{equation}
This relies on the following first-order differential relation:
\begin{equation}
    \chi\partial_\chi \left( F_h-F_{h+1} \right)=h\left( F_h+F_{h+1} \right).
    \label{xD(F-F)}
\end{equation}
Representing $F_h\left( \chi \right)$ as a series for $\chi<1$,
\begin{equation}
    F_h\left( \chi \right)=\sum_{k=0}^\infty \frac{\Gamma^2\left( h+k \right)}{\Gamma\left( 2h+k \right)\Gamma\left( k+1 \right)}\chi^{h+k},
    \label{F_series}
\end{equation}
we can show that (\ref{xD(F-F)}) indeed holds.

\section{\(SU(1,1|1)\)-invariant norm}
\label{sec:app_norm1d}

In this Section we find the $SU(1,1|1)$-invariant measure on four-point functions in terms of the $\chi$ cross-ratio.
We start with the chiral measure:
\begin{equation}
    \langle f,g \rangle = \int \frac{d\tau_1 d\bar{\theta}_1 d\tau_2 d{\theta}_2}{\langle 12 \rangle} \frac{d\tau_3 d\bar{\theta}_3 d\tau_4 d\theta_4}{\langle 34 \rangle} f^* g=\int d\mu f^* g ,
    \label{app_norm_inv}
\end{equation}
for $f,g$ satisfying (anti)chirality conditions:
\begin{equation}
    D_{1,3} f=\bar{D}_{2,4} f=D_{1,3} g=\bar{D}_{2,4} g.
    \label{fg_chiral}
\end{equation}
With the $SU(1,1|1)$ group, we can apply a superconformal transformation to all four supercoordinates. 
The infinitesimal generators of a generic transformation are:

\begin{eqnarray}    
    V_1&=& L_0^{(1)}+L_0^{(2)}+L_0^{(3)}+L_0^{(4)},\\
    V_2&=& L_1^{(1)}+L_1^{(2)}+L_1^{(3)}+L_1^{(4)},\\
    &&\cdots\\
    V_7&=& \bar{G}_{-1/2}^{(1)}+\bar{G}_{-1/2}^{(2)}+\bar{G}_{-1/2}^{(3)}+\bar{G}_{-1/2}^{(4)},
    \label{V_sum}
\end{eqnarray}
the generators being listed in the Appendix \ref{sec:app_Casimir}.
With seven generators, we can fix seven coordinates $\tau_{2,3,4}$, $\bar{\theta}_{1,3}, \theta_{2,4}$, leaving only $\tau_1=\chi$. (The final answer won't depend on $\bar{\theta}_{2,4}$ or $\theta_{1,3}$, so we are not fixing those.)
We wish to find the invariant measure as a function of $\chi$.
In other words, the group action allows us to define a map:
\begin{equation}
    \varphi: \mathbb R^{4|4} \to \mathbb R,
    \label{app_phi_def}
\end{equation}
and we are looking for the  invariant measure $d\mu\left( \chi \right)$ on $\mathbb R$ which is a pushforward of the measure $d\mu$ on $\mathbb R^{4|4}$.
This measure can be found as a contraction of the infinitesimal generators $V_i$ with the original measure $d\mu$:
\begin{equation}
    d\mu\left( \chi \right)=\left.\imath_{V_1} \imath_{V_2} \dots \imath_{V_7}\right|_{\frac{\partial}{\partial \tau_1}=0}d\mu,
    \label{dmu(chi)}
\end{equation}
with the generator of transformation along $\tau_1$ not acting, so that we can keep the $\tau_1$ coordinate.
This contraction is given by a superdeterminant:
\begin{equation}
    \left.\left.\imath_{V_1} \imath_{V_2} \dots \imath_{V_7}\right|_{\frac{\partial}{\partial \tau_1}=0}d\mu\right|_{\bar{\theta}_{1,3}= \theta_{2,4}=0}= \frac{1}{\tau_1-\tau_2} \frac{1}{\tau_3-\tau_4}\Ber\left(
    \begin{array}{ccc|cccc}
        -1 & -1 & -1 & &&&\\
        -\tau_2 & -\tau_3 & -\tau_4 & &&&\\
        -\tau_2^2 & -\tau_3^2 & -\tau_4^2 & &&&\\
        \hline
        0 & -\tau_3 \theta_3 & 0 & 0 & 0 & \tau_1 & \tau_3\\
        0 & - \theta_3 & 0 & 0 & 0 & 1 & 1\\
        -\tau_2 \bar{\theta}_2 & 0 & -\tau_4 \bar{\theta}_4 & \tau_2 & \tau_4 & 0 & 0\\
        -\bar{\theta}_2 & 0 & -\bar{\theta}_4 & 1 & 1 & 0 & 0
    \end{array}
    \right),
    \label{app_sdet}
\end{equation}
which gives:
\begin{equation}
    d\mu\left( \tau_1,\tau_2,\tau_3,\tau_4 \right)=\frac{\left( \tau_2-\tau_3 \right)\left( \tau_3-\tau_4 \right)\left( \tau_2-\tau_4 \right)}{\left( \tau_2-\tau_4 \right)\left( \tau_1-\tau_3 \right)\left( \tau_1-\tau_2 \right)\left( \tau_3-\tau_4 \right)}d\tau_1=\frac{\tau_2-\tau_3}{\left( \tau_1-\tau_3 \right)\left( \tau_1-\tau_2 \right)}d\tau_1.
    \label{dmu(tau)}
\end{equation}
Fixing further the even coordinates to be:
\begin{equation}
    \tau_1=\chi, \qquad \tau_2=0, \qquad \tau_3=1, \qquad \tau_4=\infty, 
    \label{app_tau_fix}
\end{equation}
we find:
\begin{equation}
    d\mu\left( \chi \right)=\frac{d\chi}{\chi\left( 1-\chi \right)}.
    \label{app_dmu(chi)}
\end{equation}

\section{Normalization of bound states}
\label{sec:app_norm_bound}

In this Appendix we prove the relation (\ref{norm_bound}).
To do that, we first consider the norm of non-supersymmetric SYK model.
Let's take the expression:
\begin{equation}
    \langle C_{\mathcal N=0} \Psi^A_{h'}, \Psi^A_h \rangle_0 - \langle \Psi^A_{h'}, C_{\mathcal N=0}\Psi^A_h \rangle_0.
    \label{app_Chh'}
\end{equation}
Zero subscript signifies the $\mathcal N=0$ norm:
\begin{equation}
    \langle f,g \rangle_0 = \int_{-\infty}^\infty \frac{d\chi}{\chi^2}f^*g.
    \label{app_norm_0}
\end{equation}
For distinct $h,h'$ this expression should be zero to ensure hermiticity; however if we take $h,h'$,
\begin{equation}
    h'=h+\epsilon,
    \label{h_eps}
\end{equation}
it should be proportional to $\epsilon$:
\begin{equation}
    \langle C_{\mathcal N=0} \Psi^A_{h'}, \Psi^A_h \rangle_0 - \langle \Psi^A_{h'}, C_{\mathcal N=0}\Psi^A_h \rangle_0=\epsilon \left( 2h-1 \right) \langle \Psi^A_h, \Psi^A_h \rangle_0.
    \label{app_Chh'_eps}
\end{equation}
On the other hand, using the explicit form of the Casimir (\ref{C0_C2}) and the norm (\ref{app_norm_0}), we find:
\begin{equation}
    \langle C_{\mathcal N=0} \Psi^A_{h'}, \Psi^A_h \rangle_0 - \langle \Psi^A_{h'}, C_{\mathcal N=0}\Psi^A_h \rangle_0 = \left. \Psi^A_{h'} \left( 1-\chi \right)\partial_\chi \Psi^A_h - \Psi^A_h \left( 1-\chi \right)\partial_\chi \Psi^A_{h'}\right|_{-\infty}^\infty.
    \label{app_C_inf}
\end{equation}
The eigenfunction $\Psi^A_h\left( \chi \right)$ behaves as a logarithm at infinity:
\begin{equation}
    \chi \to \infty: \qquad \Psi^A_h \sim a(h)+b(h)\log \chi+O\left( \frac{1}{\chi} \right),
    \label{app_psi_asymp}
\end{equation}
which implies that:
\begin{equation}
    \left. \Psi^A_{h'}\partial_\chi \Psi^A_h - \Psi^A_h \partial_\chi \Psi^A_{h'}\right|_{-\infty}^\infty=0.
    \label{psi_d_psi_inf}
\end{equation}
Using formula for the norm of an $\mathcal N=0$ bound state in the right-hand side of (\ref{app_Chh'_eps}),
\begin{equation}
    \langle \Psi^A_h, \Psi^A_h \rangle_0=\frac{4\pi^2}{|2h-1|},
    \label{psi_psi_0}
\end{equation}
we find the relation:
\begin{equation}
    4\pi^2 \epsilon \cdot \sgn\left( h-\frac12 \right)=\left. \Psi^A_h \chi\partial_\chi \Psi^A_{h'}-\Psi^A_{h'}\chi\partial_\chi \Psi^A_h\right|_{-\infty}^\infty, \qquad h'=h+\epsilon.
    \label{psi_asymp_eps}
\end{equation}
Luckily, this relation allows us to find the norm of the $\mathcal N=2$ eigenstates as well.
Indeed, consider two $\mathcal N=2$ eigenfunctions for close values of $h$.
By the same token as before, we have:
\begin{equation}
    \langle \mathcal C \xi_{h'}, \xi_h \rangle - \langle \xi_{h'}, \mathcal C \xi_h\rangle = 2h \epsilon \langle \xi_{h}, \xi_h \rangle = \left. \xi_{h'} \chi \partial_\chi \xi_h - \xi_h \chi \partial_\chi \xi_{h'}\right|_{-\infty}^\infty, \qquad h'=h+\epsilon.
    \label{C_xi_xi}
\end{equation}
Since the $\mathcal N=2$ eigenfunction is a linear combination of the non-supersymmetric ones,
\begin{equation}
    \xi_h=h\left( \Psi^A_h-\Psi^S_{h+1} \right), 
    \label{app_xi_psi_psi}
\end{equation}
and the non-supersymmetric functions of different types are orthogonal,
\begin{equation}
    \langle \Psi^A_h, \Psi^S_{h'}\rangle_0\equiv0,
    \label{A_S_ortho}
\end{equation}
we can rewrite (\ref{C_xi_xi}) as:
\begin{equation}
    2h \epsilon \langle \xi_{h}, \xi_h \rangle =h^2 \left.\left( \Psi^A_{h'}\chi \partial_\chi \Psi^A_h-\Psi^A_h \chi \partial_\chi \Psi^A_{h'}+\Psi^S_{h'+1}\chi \partial_\chi \Psi^S_{h+1}-\Psi^S_{h+1}\chi \partial_\chi \Psi^S_{h'+1} \right)\right|_{-\infty}^\infty.
    \label{eps_xi_psi}
\end{equation}
Using the relation we have found in the non-supersymmetric model (\ref{psi_asymp_eps}) (and an analogous relation for the $\Psi^S_h$ eigenfunctions), we finally find:
\begin{equation}
    \langle \xi_h, \xi_h \rangle = 4\pi^2 |h|.
    \label{app_xi_norm_bound}
\end{equation}

\section{Eigenvalues of the kernel}
\label{sec:app_kernel}
Let's compute the integral:

\begin{equation}
    \int K f^A\left( 1,2,\infty \right)= \frac{\tan \pi \Delta}{4\pi} \int d\tau_1 d\tau_2 d\bar{\theta}_1 d\theta_2 \frac{1 }{|\langle 12 \rangle|^{1-2\Delta - h}}\frac{\sgn \left( \tau_1'-\tau_2 \right)}{|\langle 1'2 \rangle|^{2\Delta}}\frac{\sgn \left( \tau_1-\tau_2' \right)}{|\langle 12' \rangle|^{2\Delta}},
    \label{kA_int_app}
\end{equation}
where we take three-point function in the form:
\begin{equation}
    f^A\left( 1,2,\infty \right)=\frac{\sgn \left( \tau_1-\tau_2 \right)}{|\langle 12\rangle|^{2\Delta-h}}.
    \label{fA_infty}
\end{equation}

We can fix the odd coordinates of the points $1', 2'$  to be $(\theta_1',0),(0, \bar{\theta}_2)$ and then take the Grassmann integral.
The result is:

\begin{equation}
    \int K f^A\left( 1,2,0 \right)= 2\left( -1+h+2\Delta \right)\frac{\tan \pi \Delta}{4\pi} \int d\tau_1 d\tau_2  \frac{\sgn \left( \tau_2-\tau_1 \right) }{| \tau_1-\tau_2 |^{2-2\Delta - h}}\frac{\sgn \left( \tau_1'-\tau_2 \right)}{|\tau_1'-\tau_2|^{2\Delta}}\frac{\sgn \left( \tau_1-\tau_2' \right)}{|\tau_1-\tau_2'|^{2\Delta}}. 
    \label{kA_int_app_1}
\end{equation}
Changing variables:
\begin{eqnarray}
    \tau_1&=& \left( \tau_2'-\tau_1' \right)v+\tau_1',\\
    \tau_2&=& \left( \tau_2'-\tau_1' \right)u+\tau_1',
    \label{change1}
\end{eqnarray}
we see that the anti-symmetric three-point function is indeed an eigenvector of the kernel:

\begin{equation}
    \int K f^A\left( 1,2,0 \right)=\frac{\sgn \left( \tau_1'-\tau_2' \right)}{|\tau_1'-\tau_2'|^{2\Delta-h}}\cdot k^A,
    \label{K_diag}
\end{equation}
where the eigenvalue is:
\begin{equation}
    k^A=2\left( -1+h+2\Delta \right)\frac{\tan \pi \Delta}{4\pi} \int dudv \frac{\sgn\left( u-v \right)\sgn \left( 1-v \right) \sgn u}{|u-v|^{2-2\Delta-h}|u|^{2\Delta} |v-1|^{2\Delta}}. 
    \label{kA_uv}
\end{equation}
Changing variables further:
\begin{equation}
    u=vw,
    \label{change2}
\end{equation}
we see that the integral splits into two of the same type:
\begin{equation}
    k^A=-2\left( -1+h+2\Delta \right)\frac{\tan \pi \Delta}{4\pi} \int dv\frac{\sgn v\sgn\left( v-1 \right)}{|v|^{1-h} |v-1|^{2\Delta}} \int dw\frac{\sgn w\sgn\left( w-1 \right)}{|w|^{2\Delta} |w-1|^{2-2\Delta-h}}. 
    \label{ka_3}
\end{equation}
Using the integral definition of the beta-function, we find:
\begin{equation}
    \int dt \frac{\sgn t \sgn \left( t-1 \right)}{|t|^a |t-1|^b}=B\left( 1-a,-1+a+b \right)-B\left( 1-a,1-b \right)+B\left( 1-b,-1+a+b \right).
    \label{int_beta}
\end{equation}
Using various identities, we arrive at the answer (\ref{kA_ans}).
The symmetric eigenvalue is recovered from $h \leftrightarrow -h$ symmetry:

\begin{equation}
    k^S\left( h \right)=k^A\left( -h \right).
    \label{kS_app}
\end{equation}

\section{Zero-rung propagator}
\label{sec:app_0rung}

In this Appendix, we find the inner product of an eigenfunction with a zero-rung propagator:
\begin{equation}
    \langle \xi_h(\chi), \chi^{2\Delta}\rangle.
    \label{app_proj_def}
\end{equation}
As before, it is instructive to consider the same problem in the non-supersymmetric model.
Let's denote the corresponding product by $n_0\left( h,\Delta \right)$:
\begin{equation}
    n^A_0\left( h,\Delta \right)\equiv \langle \Psi^A_h, \chi^{2\Delta} \rangle_0=\frac12\alpha_0 k_0^A(h), \qquad \alpha_0=\frac{2\pi \Delta}{\left( 1-\Delta \right)\left( 1-2\Delta \right)}\cot \pi \Delta.
    \label{n0_def}
\end{equation}
Applying the Casimir to the functions inside the product and using the hermiticity, we find:
\begin{equation}
   \langle \mathcal C_{\mathcal N=0} \Psi^A_h,\chi^{2\Delta}\rangle_0=h(h-1)\langle \Psi^A_h,\chi^{2\Delta}\rangle_0=\langle\Psi^A_h, \mathcal C_{\mathcal N=0} \chi^{2\Delta}\rangle_0=2\Delta\left( 2\Delta-1 \right)\langle \Psi^A_h,\chi^{2\Delta}\rangle_0-4\Delta^2\langle \Psi^A_h,\chi^{2\Delta+1}\rangle_0.
    \label{C_psi_chi}
\end{equation}
This gives us the following identity:
\begin{equation}
    n_0^A\left( h, \Delta+\frac12 \right)=\frac{\left( 2\Delta-h \right)\left( 2\Delta+h-1 \right)}{4\Delta^2}n_0^A\left( h,\Delta \right).
    \label{n+1/2}
\end{equation}
Now we can follow the same line of reasoning for the $\mathcal N=2$ eigenfunctions. 
Acting with the Casimir on the inner product \ref{app_proj_def}, we get:
\begin{equation}
    \langle \mathcal C \xi_h, \chi^{2\Delta}\rangle=h^2 \langle \xi_h, \chi^{2\Delta}\rangle=\langle \xi_h, \mathcal C \chi^{2\Delta}\rangle=4\Delta^2 \langle \xi_h, \chi^{2\Delta+1}\rangle_0.
    \label{C_xi_chi}
\end{equation}
Using again the relation (\ref{Xi_def}) between $\mathcal N=0$ and $\mathcal N=2$ eigenfunctions, we find:
\begin{equation}
    \langle \xi_h, \chi^{2\Delta}\rangle=\frac{4\Delta^2}{h}\left( n_0^A\left( h,\Delta+\frac12 \right)-n_0^S\left( h+1,\Delta+\frac12 \right) \right).
    \label{xi_chi_n}
\end{equation}
We need two more identities: the relation between symmetric and antisymmetric eigenvalues (following from (\ref{kA_ans_n0}, \ref{kB_ans_n0}),
\begin{equation}
    \frac{k_0^S\left( h+1,\Delta+\frac12 \right)}{k_0^A\left( h,\Delta+\frac12 \right)}=\frac{2\Delta+h}{2\Delta-h},
    \label{kS_kA}
\end{equation}
and the relation between $\mathcal N=0$ and $\mathcal N=2$ eigenvalues (\ref{k_n2_n0}):
\begin{equation}
    k^A\left( h, \Delta \right)=\frac{2\Delta +h-1}{2\Delta-2}k_0^A\left( h,\Delta \right).
    \label{app_k2_k0}
\end{equation}
Bringing together (\ref{n+1/2}, \ref{xi_chi_n}, \ref{kS_kA}, \ref{app_k2_k0}), we finally find:
\begin{equation}
    \langle \xi_h, \chi^{2\Delta}\rangle=\frac12\alpha k^A\left( h \right).
    \label{app_0rung_k}
\end{equation}
\bibliographystyle{JHEP}

\bibliography{n=2}
\end{document}